\newif\ifpreprint

\preprintfalse 

\ifpreprint
\documentclass[journal=jctcce,manuscript=letter]{achemso}
\else
\documentclass[journal=jctcce,manuscript=letter,layout=twocolumn]{achemso}
\fi

\usepackage[T1]{fontenc} 

\usepackage{amsmath}
\usepackage{newtxtext,newtxmath}

\usepackage{graphicx}
\usepackage{dcolumn}
\usepackage{braket}
\usepackage{longtable}
\usepackage{multirow}
\usepackage{threeparttable}
\usepackage{xspace}
\usepackage{verbatim}
\usepackage[version=4]{mhchem} 
\usepackage{comment}
\usepackage{color,soul}

\usepackage{mathtools}
\usepackage[dvipsnames]{xcolor}
\usepackage{xspace}
\usepackage{ifthen}

\usepackage{qcircuit}

\usepackage{graphicx,longtable,dcolumn,mhchem}
\usepackage{rotating,color}
\usepackage{lscape}
\usepackage{amsmath}
\usepackage{dsfont}
\usepackage{soul}
\usepackage{physics}
\newcolumntype{d}{D{.}{.}{-1}}

\newcommand{\ra}{\rightarrow}
\newcommand{\pis}{\pi^\star}
\newcommand{\npi}{$n \rightarrow \pi^\star$}
\newcommand{\ppi}{$\pi \rightarrow \pi^\star$}

\newcommand{\Val}{\mathrm{V}}

\newcommand{\si}{\sigma}

\newcommand{\AVTZ}{\emph{aug}-cc-pVTZ}

\usepackage[colorlinks = true,
            linkcolor = blue,
            urlcolor  = black,
            citecolor = blue,
            anchorcolor = black]{hyperref}

\definecolor{goodorange}{RGB}{225,125,0}
\definecolor{goodgreen}{RGB}{5,130,5}
\definecolor{goodred}{RGB}{220,50,25}
\definecolor{goodblue}{RGB}{30,144,255}

\newcommand{\note}[2]{
\ifthenelse{\equal{#1}{F}}{
\colorbox{goodorange}{\textcolor{white}{\footnotesize \fontfamily{phv}\selectfont #1}}
    \textcolor{goodorange}{{\footnotesize \fontfamily{phv}\selectfont #2}}\xspace
}{}
\ifthenelse{\equal{#1}{R}}{
\colorbox{goodred}{\textcolor{white}{\footnotesize \fontfamily{phv}\selectfont #1}}
    \textcolor{goodred}{{\footnotesize \fontfamily{phv}\selectfont #2}}\xspace
}{}
\ifthenelse{\equal{#1}{N}}{
\colorbox{goodgreen}{\textcolor{white}{\footnotesize \fontfamily{phv}\selectfont #1}}
    \textcolor{goodgreen}{{\footnotesize \fontfamily{phv}\selectfont #2}}\xspace
}{}
\ifthenelse{\equal{#1}{M}}{
\colorbox{goodblue}{\textcolor{white}{\footnotesize \fontfamily{phv}\selectfont #1}}
    \textcolor{goodblue}{{\footnotesize \fontfamily{phv}\selectfont #2}}\xspace
}{}
}

\usepackage{titlesec}

\usepackage[fontsize=11pt]{scrextend}
\titleformat{\section}
{\normalfont\sffamily\bfseries\color{Blue}}
{\thesection.}{0.25em}{\uppercase}

\titleformat{\subsection}
{\normalfont\sffamily\bfseries}
{\thesubsection}{0.25em}{}

\titleformat{\subsubsection}
{\normalfont\sffamily}
{\thesubsubsection}{0.25em}{}

\titleformat{\suppinfo}
{\normalfont\sffamily\bfseries}
{\thesubsection}{0.25em}{}

\titlespacing*{\section}{0pt}{0.5\baselineskip}{0.01\baselineskip}
\titlespacing*{\subsection}{0pt}{0.125\baselineskip}{0.01\baselineskip}
\titlespacing*{\subsubsection}{0pt}{0.125\baselineskip}{0.01\baselineskip}

\author{Rudraditya Sarkar}
	\affiliation[CEISAM, Nantes]{Universit\'e de Nantes, CNRS,  CEISAM UMR 6230, F-44000 Nantes, France}
\author{Pierre-Fran\c{c}ois Loos}
	\affiliation[LCPQ, Toulouse]{Laboratoire de Chimie et Physique Quantiques, Universit\'e de Toulouse, CNRS, UPS, France}
\author{Martial Boggio-Pasqua}
	\email{martial.boggio@irsamc.ups-tlse.fr}
	\affiliation[LCPQ, Toulouse]{Laboratoire de Chimie et Physique Quantiques, Universit\'e de Toulouse, CNRS, UPS, France}
\author{Denis Jacquemin}
	\email{Denis.Jacquemin@univ-nantes.fr}
	\affiliation[CEISAM, Nantes]{Universit\'e de Nantes, CNRS,  CEISAM UMR 6230, F-44000 Nantes, France}
		
\setkeys{acs}{usetitle=true}

\let\oldmaketitle\maketitle
\let\maketitle\relax
     \title{Assessing the Performances of CASPT2 and NEVPT2 for Vertical Excitation Energies}
\date{\today}

\begin{tocentry}
	\centering
	\includegraphics[width=1.\textwidth]{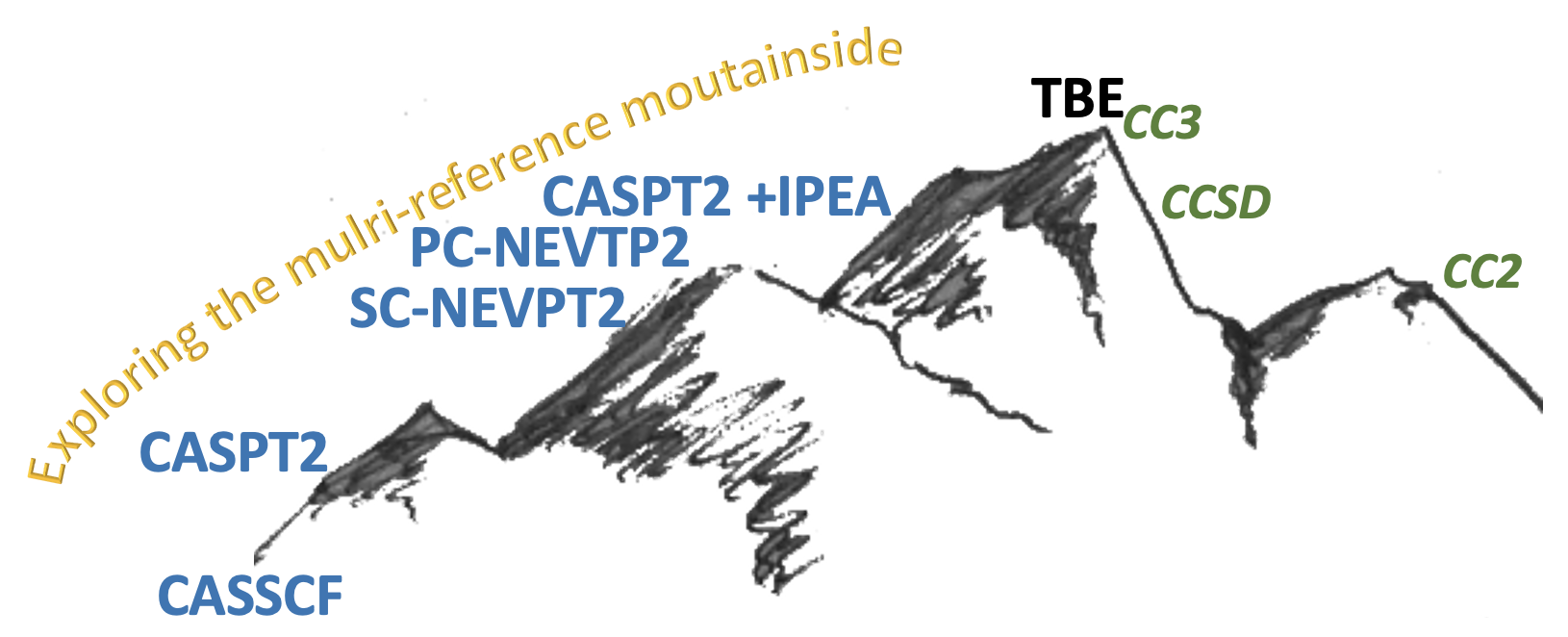}
\end{tocentry}

\begin{document}

\ifpreprint
\else
\twocolumn[
\begin{@twocolumnfalse}
\fi
\oldmaketitle

\begin{abstract}
{Methods able to simultaneously account for both static and dynamic electron correlations have often been employed, not only to model photochemical events, but also to provide reference values 
for vertical transition energies, hence allowing to benchmark lower-order models.} In this category, both the complete-active-space second-order perturbation theory (CASPT2) and the $N$-electron valence state second-order
perturbation theory (NEVPT2) are certainly popular, the latter presenting the advantage of not requiring the application of the empirical 
ionization-potential-electron-affinity (IPEA) and level shifts. However, the actual accuracy of these multiconfigurational approaches is not settled yet. {In this context,  to assess the performances of these approaches
the present work relies on highly-accurate ($\pm$ 0.03 eV) \emph{aug}-cc-pVTZ vertical transition energies for 284 excited states of diverse character} (174 singlet, 110 triplet, 206 valence, 78 Rydberg, 78 \npi, 119 \ppi, 
and 9 double excitations) determined in 35 small- to medium-sized organic molecules containing from three to six non-hydrogen atoms. 
The CASPT2 calculations are performed with and without IPEA shift and compared to the partially-contracted (PC) and strongly-contracted (SC) variants of NEVPT2. We find that both CASPT2 with IPEA shift 
and PC-NEVPT2 provide fairly reliable vertical transition energy estimates, with slight overestimations and mean absolute errors of $0.11$ and $0.13$ eV, respectively. These values are found to be 
rather uniform for the various subgroups of transitions. The present work completes our previous benchmarks focussed on single-reference wave function methods (\textit{J.~Chem. Theory Comput.} 
\textbf{14}, 4360 (2018); \emph{ibid.}, \textbf{16}, 1711 (2020)), hence allowing for a fair comparison between various families of electronic structure methods. 
In particular, we show that ADC(2), CCSD, and CASPT2 deliver similar accuracies for excited states with a dominant single-excitation character.
\end{abstract}

\ifpreprint
\else
\end{@twocolumnfalse}
]
\fi

\ifpreprint
\else
\small
\fi

\noindent

\section{Introduction}

In the early 90's, it was certainly challenging to systematically introduce dynamical electron correlation effects in the description of molecular systems, so as to access accurate geometries, dissociation energies, vibrational 
frequencies, spectroscopic constants, as well as vertical transition energies (VTEs). In this framework, the development and efficient implementation of single-reference many-body perturbation theory (SR-MBPT) based on 
a Hartree-Fock (HF) reference wave function were certainly major steps forward for the electronic structure community. At that time, most applications using SR-MBPT were based on the acclaimed second-order M{\o}ller-Plesset 
perturbative correction, \cite{Mol34} and while the reliability of such approach was found to be quite satisfactory for closed-shell systems, it obviously suffers from major drawbacks when the HF reference wave function is not a 
valid starting point, i.e., when the targeted electronic state is not properly described by a single Slater determinant. This situation is ubiquitous, for example, in transition metal complexes, bond-breaking reactions, transition states, 
and, more importantly in the present context, electronic excited states (ESs). This severe limitation has led to the development of multireference extensions of MBPT (MR-MBPT) in which the reference wave function is 
multiconfigurational, i.e., contains more than one Slater determinant. These powerful approaches allow to account for both static electron correlation encapsulated in the reference multiconfigurational wave function --- most often 
described within the complete-active-space self-consistent field (CASSCF) formalism --- and dynamic electron correlation treated very efficiently at low order of perturbation theory. 

Various flavors and implementations of MR-MBPT have emerged over the years. Among the most well-known variants, one can cite the complete-active-space second-order perturbation theory (CASPT2) developed by Roos and 
coworkers, \cite{And90,And92} the multireference second-order M{\o}llet-Plesset (MRMP2) approach proposed by Hirao, \cite{Hir92} and the $N$-electron valence state second-order perturbation theory (NEVPT2) designed by 
Angeli, Malrieu, and coworkers. \cite{Ang01,Ang01b,Ang02} All these methods have proven to be efficient and accurate when one deals with ESs, in particular for the computation of VTEs. Illustratively, three decades after its original formulation, 
CASPT2 has clearly become the most popular multiconfigurational approach of molecular quantum chemistry in order to tackle ground states and ESs of multiconfigurational character on an equal footing.

Nevertheless, it is noteworthy that several limitations of CASPT2 were rapidly disclosed and circumvented. The first recognized drawback of CASPT2 in the context of ESs was the appearance of intruder states, i.e., perturbers 
that have nearly equal energies to the zeroth-order CASSCF wave function. \cite{And94,And95,Roo95b} The best remedy to this severe issue is to move the intruder states in the active space but this strategy is obviously intrinsically 
limited by the dreadful computational cost associated with the expansion of the active space. An efficient alternative was formulated by Roos and Andersson \cite{Roo95} who proposed to introduce a real-valued shift in the energy 
denominators of the second-order energy (thus avoiding singularities) and correcting accordingly the resulting energy. This approach, known as \emph{level-shift correction}, is almost systematically applied nowadays. Note that, in 
the case of ``weak'' intruder states, an imaginary shift can be also introduced. \cite{For97b}

The second bottleneck of the single-state (also called state-specific) CASPT2 (SS-CASPT2) method, as implemented originally by Roos and coworkers, is the difficulty to deal with electronic state mixing (e.g., valence/Rydberg and 
covalent/ionic mixing). A solution consists in allowing the states to mix within the MR-MBPT treatment, giving rise to the multistate formulation of CASPT2 denoted as MS-CASPT2, \cite{Fin98} and later refined in an extended 
multistate version called XMS-CASPT2. \cite{Shi11b} Very recently, a further extension known as extended dynamically weighted CASPT2 (XDW-CASPT2) was proposed in order to combine the most attractive features of both 
MS-CASPT2 and XMS-CASPT2. \cite{Bat20}

The third bottleneck was found in evaluating a large number of chemical problems for which systematic errors were noticed  \cite{And93b,And95b}  and ascribed to the unbalanced description of the zeroth-order Hamiltonian 
for the open- and closed-shell electronic configurations. This systematic error can be attenuated by introducing an additional parameter, the so-called ionization-potential-electron-affinity (IPEA) shift, in the zeroth-order 
Hamiltonian. \cite{Ghi04} 

Whilst these three ``fixes'' improve the overall accuracy of CASPT2, all of them remain rather empirical, which can be viewed as a drawback compared to other \emph{ab initio} approaches. In this context, NEVPT2 has the advantage 
of being practically free of intruder states and enjoys the valuable property of being size-consistent (which is not the case of CASPT2). NEVPT2 mainly differs from CASPT2 in the choice of (i) the zeroth-order Hamiltonian, and (ii) the 
definition of the perturbers and their corresponding energies. NEVPT2 exists in two different contraction schemes: strongly-contracted (SC) and partially-contracted (PC), the latter being supposedly more accurate thanks to its larger 
number of perturbers and its greater flexibility. Note that a multistate version of NEVPT2 known as quasi-degenerate NEVPT2 (QD-NEVPT2) has also been developed. \cite{Ang06}

Both the parent and stepwisely improved versions of CASPT2 have been applied to tackle a wide variety of chemical problems: heavy element chemistry, \cite{Gag07} biochemical systems, \cite{Sie03b,Sch10b} transition metal 
complexes \cite{Fou05,Pie06b,Nee06,Ord08,Sua09,Ban08,Kep09} (such as bimetallic complexes \cite{Que08}), and, of course, ESs. The latter were intensively investigated by Roos, Serrano-Andr\'es, and their collaborators from the start,
\cite{Ser93,Ser93b,Ser93c,Ser95,Roo96,Ser96,Ser96b,Ser98b,Roo99,Mer99,Roo02,Ser02,Ser05} in works where experiment was typically used to appraise the quality of the CASPT2 estimates. These estimates were themselves used later 
as references in benchmark studies assessing the performances of ``lower-level'' ES methods. \cite{Toz99,Bur02,Pea08,Fab13} Starting in 2008, very comprehensive benchmarks of valence VTEs in small and medium CNOH compounds  
were performed by Thiel's group. \cite{Sch08,Sil08,Sau09,Sil10,Sil10b,Sil10c} Besides literature data, these authors first relied on CASPT2/TZVP to define a list of 104 theoretically best estimates (TBEs) for the singlet vertical transitions in 28 organic 
molecules optimized at the MP2/6-31G(d) level. In their original study, \cite{Sch08} the TBEs associated with 63 triplet ESs were obtained at the third-order coupled-cluster (CC3) \cite{Chr95b} level with the same TZVP basis set. These different choices
of reference methods for the two ES families were justified by the almost pure single-excitation character of the triplet transitions, whereas the singlet transitions showed a less clear-cut nature. In 2010, Thiel's group upgraded their 
TBEs using the \emph{aug}-cc-pVTZ basis set \cite{Sil10b,Sil10c} and found that using this larger basis set containing additional diffuse functions downshifted the singlet VTEs by an average of $0.11$ eV as compared to TZVP, with a high 
degree of consistency between the two sets of data (correlation coefficient larger than $0.996$). \cite{Sil10b}   Interestingly, in their studies employing the \emph{aug}-cc-pVTZ basis set, Thiel and coworkers also shifted to 
CC3 as the default reference approach for defining the TBEs associated with singlet transitions. Yet, no error bar was defined, nor estimated for CASPT2.  

Both the  original TZVP and the improved  \emph{aug}-cc-pVTZ  TBEs of Thiel and coworkers were later employed in countless benchmarks of ES models.
\cite{Sil08,Goe09,Jac09c,Roh09,Sau09,Jac10c,Jac10g,Sil10,Mar11,Jac11a,Hui11,Del11,Tra11,Pev12,Dom13,Dem13,Sch13b,Voi14,Har14,Yan14b,Sau15,Pie15,Taj16,Mai16,Ris17,Dut18,Hel19}
In particular, we wish to pinpoint valuable studies of NEVPT2/CASPT2 \cite{Sch13b} and CASSCF \cite{Hel19} using Thiel's TZVP reference values. These works are clearly in the same philosophy as the present
effort, though these were obviously limited to valence transitions of single-excitation nature.
At the same time, the number of ES studies employing CASPT2 steadily increased during the last two decades, with investigations of many photophysical and photochemical processes: photosensitization, photoisomerization, 
charge transfer, nonradiative deactivation, as well as photodynamics studies of small to medium size organic molecules. 
\cite{Roc12,Seg12,Li12d,Gob12,Gob12b,Guo12b,Yam12,Goz12b,Mel12,Gad12b,Gob13,Fan13,Cho13,Hui13,Nak13,Kom13,Per13b,Giu14,Zob15,Mai15b,Nen15,Elz15,Dum15,Mat15,Seg16,Bao17,Giu17,Hei19,Giu19,Avi19}

Given such extensive use of CASPT2 by both the electronic structure and dynamics communities, Gonz\'{a}lez and co-workers proposed an excellent reassessment of this approach in 2017. \cite{Zob17} In this key work, they showed that,
for di- and tri-atomic molecules for which full configuration interaction (FCI) energies could be computed (albeit with a small basis set), standard CASPT2 \cite{And92} slightly underestimates the VTEs, while the application of the IPEA shift \cite{Ghi04} 
partially corrects this underestimation. The same work also reports that applying an IPEA shift can lead to overestimations of the VTEs for medium-sized molecules, and even quite significant exaggerations for larger 
molecules (of the same order of magnitude as the underestimation obtained without IPEA for the smallest derivatives). As a result, the authors of Ref.~\citenum{Zob17} concluded that the application of the IPEA shift is not systematically justified 
for organic chromophores. However, and importantly, this work also shows that the relevance of applying IPEA is basis set dependent: for double-$\zeta$ basis sets, smaller errors are obtained when setting the IPEA shift to zero, whereas 
for more extended basis set, of triple- or quadruple-$\zeta$ quality, applying an IPEA shift improves the accuracy. In such context, one can find several other examples in the literature showing the importance of introducing an IPEA shift
to reach a better agreement between experiment and theory, as for, e.g., the excitation energies in iron complexes, \cite{Pie06b,Pie08,Sua09,Kep09,Dak12,Rud14,Vel16} or in BODIPY derivatives. \cite{Wen18}

Of course, besides CASPT2 and related MR-MBPT, there exist many alternative \emph{ab initio} methods for tackling ESs. In this regard, one should certainly cite several efficient single-reference approaches for modeling VTEs in large systems: 
(i) time-dependent density-functional theory (TD-DFT), \cite{Cas12,Ulr12b} (ii) the Bethe-Salpeter equation (BSE) formalism, \cite{Bla18,Bla20} (iii) the second-order algebraic-diagrammatic construction [ADC(2)], \cite{Dre15} and (iv) the second-order 
coupled-cluster (CC2) model. \cite{Chr95,Hat00}  However, the overall accuracy provided by these four methodologies is typically  insufficient to be considered as a safe reference for assessing other methods. \cite{Loo20c}  We underline here that
in contrast to TD-DFT and BSE, both the ADC and CC advantageously offer a path for systematic improvement through the increase of the expansion order, with, e.g., third-order approaches like ADC(3) \cite{Har14} and CC3. \cite{Chr95b}  
In this context, some of us, inspired by Thiel's works,  have recently proposed a new set of highly-accurate TBEs for VTEs of a large variety of ESs in small- and medium-sized organic molecules and radicals. 
\cite{Loo18a,Loo19c,Loo20a,Loo20b,Loo20d,Ver21,Chr21,Loo21a} 
These TBEs were obtained directly from FCI using a selected CI approach, \cite{Gar18,Sce19} CCSDTQ,  \cite{Kuc91,Kal03,Kal04,Hir04} and CCSDT, \cite{Nog87,Scu88,Kuc01,Kow01,Kow01b} for compounds containing 
1-to-3, 4, and 5-to-10 non-hydrogen atoms, respectively. Taking advantages of these highly trustworthy values, we were able to resolve some unanswered questions about the relative accuracies of ADC(3),  \cite{Har14} CCSDT-3, \cite{Wat96,Pro10} and CC3. 
\cite{Chr95b}  In particular, we evidenced that the accuracy of the VTEs obtained with CCSDTQ is comparable to that of FCI, whereas both CC3 and CCSDT-3 can be viewed as highly reliable, at least for transitions with a predominant 
single-excitation character.  \cite{Loo18a,Loo19c,Loo20a,Loo20b,Loo20d,Ver21,Chr21}

\begin{figure}[htp]
\centering
 \includegraphics[width=\linewidth,viewport=1.cm 10cm 18cm 27cm,clip]{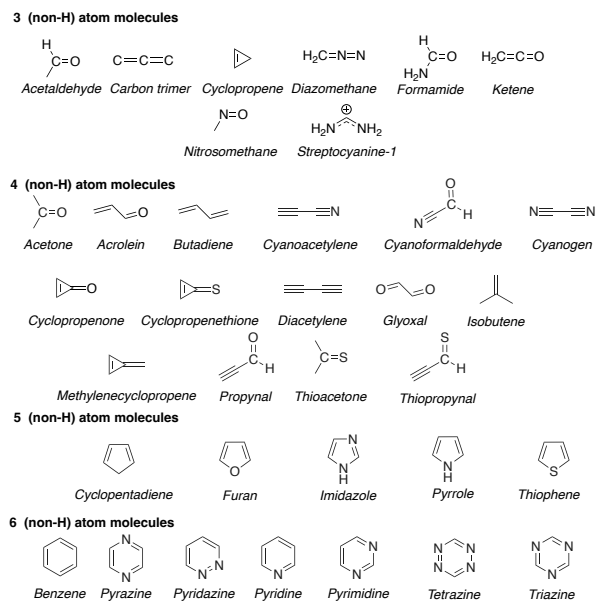}
  \caption{Various families of compounds considered in this study.}
   \label{Fig-1}
\end{figure}

In the present study, considering the ESs studied in Refs.~\citenum{Loo18a} and \citenum{Loo20a} which constitutes a set of 284 ESs (174 singlet, 110 triplet, 206 valence, 78 Rydberg, 78 \npi, 119 \ppi, and 8 double excitations) in 35 organic 
molecules containing from three to six non-hydrogen atoms (see Figure \ref{Fig-1}), we perform a comprehensive benchmark of both CASPT2 and NEVPT2, the former being considered with and without IPEA shift.  
Because we rely on very high-quality reference values, we believe that the present study can provide definite answers to the question of the relative accuracy of these multireference approaches for VTEs dominated by a 
single-excitation character, as well as reliable and trustworthy comparisons with single-reference ES methods.

\section{Computational details}

We have computed VTEs for ESs of each compound displayed in Figure \ref{Fig-1} using both {single-state} CASPT2 \cite{And90,And92,Roo95,Ghi04} and NEVPT2,  \cite{Ang01,Ang01b,Ang02} with Dunning's \emph{aug}-cc-pVTZ 
basis set. \cite{Ken92} Note that while all CASPT2 calculations are new, several (but not all) PC-NEVPT2 data were taken from a previous work. \cite{Loo20a}  The cartesian coordinates of the ground-state geometries of all 
considered molecules are taken from our previous studies. \cite{Loo18a,Loo19c,Loo20a} These high-quality ground-state equilibrium geometries have been computed at the CC3 \cite{Chr95b} level, several of them being extracted 
from earlier works. \cite{Bud17,Jac18a,Bre18a}. 

All perturbative calculations reported herein were performed on top of state-averaged (SA) CASSCF wave functions, which were produced by considering the ground state and (at least) the ES of interest. Sometimes, more 
ESs were included in the state-averaging procedure to guarantee convergence and avoid root-flipping problems. The relevant information concerning the construction of the SA-CASSCF wave functions can be found in the 
Supporting Information (SI) of the present article for all electronic states not reported in Refs.~\citenum{Loo19c} and ~\citenum{Loo20a}, whereas the corresponding information for all other electronic states can be found in 
the SI of these references.

We tackled the intruder state problem both by increasing the size of active spaces as well as by introducing a level shift ($0.3$ a.u.~unless otherwise stated) as discussed in Ref.~\citenum{Roo95}. One set of CASPT2 
calculations was performed by introducing  the standard IPEA parameter ($0.25$ a.u.) [defined below as CASPT2(IPEA)] as discussed in Ref.~\citenum{Ghi04} and another set of calculations was performed without IPEA 
correction [defined below as CASPT2(NOIPEA)]. We have performed both SC-NEVPT2 and PC-NEVPT2 calculations, and report only the latter in the main text below. The interested reader can find SC-NEVPT2 results in 
the SI. All the SA-CASSCF, CASPT2 (RS2 contraction level) \cite{molpro} and NEVPT2 calculations were performed using the MOLPRO program. \cite{MolPro19}

In the statistical analysis presented below, we report the usual statistical indicators: the mean signed error (MSE), the mean absolute error (MAE), the root-mean-square error (RMSE), the standard 
deviation of the errors (SDE), as well as largest positive and negative deviations [Max($+$) and Max($-$), respectively].

\section{Result and Discussion}

Below, we first discuss individual molecules, trying to pinpoint relevant examples, before moving to statistical analyses.   We note here that, as general trends will emerge during the discussion, we have particularly detailed the discussion on the first compounds
and tried not  to repeat general statements for each of the molecules regarding, e.g., relative accuracies for Rydberg and valence transitions, or the impact of IPEA. The full list of results can be found in Tables \ref{Table-1}, \ref{Table-2} and 
\ref{Table-3} whereas SA-CASSCF and SC-NEVPT2 results can be found in Section S3 of the SI. Except when specifically discussed below, all the ESs considered in these Tables present a strongly dominant single-excitation character.

\begin{table*}[htp]
\caption{Comparisons between TBEs taken from Table 6 of Ref.~\citenum{Loo18a}, Table 1 of Ref.~\citenum{Loo19c}, and Table 11 of Ref.~\citenum{Loo20a} and VTEs computed at the 
 CASPT2(IPEA), CASPT2(NOIPEA), and (PC-)NEVPT2 levels. $^a$} 
\label{Table-1}
\vspace{-0.4 cm}
\scriptsize
\begin{tabular}{llcccc|llcccc}
ine 
	Compound & State	& TBE & \multicolumn{2}{c}{CASPT2}    & NEVPT2 &Compound & State	& TBE & \multicolumn{2}{c}{CASPT2}    & NEVPT2 \\
	         &              &     & IPEA     & NOIPEA    &        &         &              &     & IPEA     & NOIPEA    &        \\
ine
Acetaldehyde         	&$^1A^{''} (\Val; n \ra \pis)$		        & 4.31	&4.35	&4.13	&4.39	&Cyclopropenone	&$^1B_1 (\Val; n \ra \pis)$				& 4.26 	&4.12	&3.75	&4.04 	\\
				&$^3A^{''} (\Val; n \ra \pis)$		        & 3.97	&3.94	&3.71	&4.00	&				&$^1A_2 (\Val; n \ra \pis)$				& 5.55	&5.62	&5.31	&5.85  	\\
Acetone	         	&$^1A_2 (\Val; n \ra \pis)$		      		& 4.47	&4.44	&4.19	&4.48	&				&$^1B_2 (\mathrm{R}; n \ra 3s)$		& 6.34	&6.28	&6.21	&6.51  	\\
				&$^1B_2 (\mathrm{R}; n \ra 3s)$		& 6.46	&6.46	&6.35	&6.81	&				&$^1B_2 (\Val; \pi \ra \pis)$			& 6.54	&6.54	&6.20	&6.82  	\\
				&$^1A_2 (\mathrm{R}; n \ra 3p)$	        & 7.47	&7.80	&7.55	&7.65	&				&$^1B_2 (\mathrm{R}; n \ra 3p)$		& 6.98	&6.84	&6.70	&7.07  	\\
				&$^1A_1 (\mathrm{R}; n \ra 3p)$	        & 7.51	&7.67	&7.46	&7.75	&				&$^1A_1 (\mathrm{R}; n \ra 3p)$		& 7.02	&7.27	&7.03	&7.28  	 \\
				&$^1B_2 (\mathrm{R}; n \ra 3p)$	        & 7.62	&7.56	&7.47	&7.91	&				&$^1A_1 (\Val; \pi \ra \pis)$			& 8.28	&8.96	&8.11	&8.19  	 \\
				&$^3A_2 (\Val; n \ra \pis)$		        		& 4.13	&4.13	&3.89	&4.20	&				&$^3B_1 (\Val; n \ra \pis)$				& 3.93	&3.65	&3.28	&3.51  	 \\
				&$^3A_1 (\Val; \pi \ra \pis)$	                	& 6.25	&6.24	&6.07	&6.28	&				&$^3B_2 (\Val; \pi \ra \pis)$			& 4.88	&4.76	&4.60	&5.10  	 \\
Acrolein			&$^1A'' (\Val; n \ra \pis)$		        		& 3.78	&3.70	&3.50$^c$&3.76 	&				&$^3A_2 (\Val; n \ra \pis)$				& 5.35	&5.36	&5.06	&5.60	 \\
				&$^1A' (\Val; \pi \ra \pis)$	                		& 6.69	&6.93	&6.37	&6.67 	&				&$^3A_1 (\Val; \pi \ra \pis)$			& 6.79	&6.93	&6.61	&7.16	 \\
				&$^1A'' (\Val; n \ra \pis)$		& \emph{{6.72}}  	&7.12	&6.53$^c$&7.16 	&Cyclopropenethione&$^1A_2 (\Val; n \ra \pis)$			& 3.41	&3.43	&3.14	&3.52	 \\
				&$^1A' (\mathrm{R}; n \ra 3s)$		        & 7.08	&7.12	&6.90	&7.05 	&				&$^1B_1 (\Val; n \ra \pis)$				& 3.45	&3.45	&3.17	&3.50	 \\
				&$^1A' (\Val; \pi \ra \pis)$$^b$           		& 7.87	&8.18	&7.84	&8.05	&				&$^1B_2 (\Val; \pi \ra \pis)$			& 4.60	&4.64	&4.35	&4.77	\\
				&$^3A'' (\Val; n \ra \pis)$		        		&3.51	&3.42	&3.21$^c$&3.46	&				&$^1B_2 (\mathrm{R}; n \ra 3s)$		& 5.34	&5.25	&5.15	&5.35	\\
				&$^3A' (\Val; \pi \ra \pis)$		        		& 3.94	&4.02	&3.78	&3.95 	&				&$^1A_1 (\Val; \pi \ra \pis)$			& 5.46	&5.84	&5.32	&5.54	 \\
				&$^3A' (\Val; \pi \ra \pis)$		        		& 6.18	&6.29	&5.93	&6.23 	&				&$^1B_2 (\mathrm{R}; n \ra 3p)$		& 5.92	&5.93	&5.86	&5.99	 \\
				&$^3A'' (\Val; n \ra \pis)$		& \emph{{6.54}} 	&6.81	&6.22$^c$&6.83	&				&$^3A_2 (\Val; n \ra \pis)$				& 3.28	&3.28	&3.00	&3.38	\\
Benzene			&$^1B_{2u} (\Val; \pi \ra \pis)$	        		& 5.06	&5.14	&4.66	&5.32	&				&$^3B_1 (\Val; n \ra \pis)$				& 3.32	&3.35	&3.07	&3.40 	 \\
				&$^1B_{1u} (\Val; \pi \ra \pis)$	       		& 6.45	&6.65	&6.23	&6.43	&				&$^3B_2 (\Val; \pi \ra \pis)$			& 4.01	&3.97	&3.75	&4.17 	 \\
				&$^1E_{1g} (\mathrm{R}; \pi \ra 3s)$	        & 6.52	&6.70	&6.57	&6.75	&				&$^3A_1 (\Val; \pi \ra \pis)$			& 4.01	&4.01	&3.77	&4.13 	 \\
				&$^1A_{2u}  (\mathrm{R}; \pi \ra 3p)$	& 7.08	&7.21	&7.07	&7.40	&Diacetylene		&$^1\Sigma_u^- (\Val; \pi \ra \pis)$		& 5.33	&5.42	&5.01	&5.33 	 \\
				&$^1E_{2u}  (\mathrm{R}; \pi \ra 3p)$    	& 7.15	&7.26	&7.12	&7.45	&				&$^1\Delta_u 	(\Val; \pi \ra \pis)$		& 5.61	&5.68	&5.30	&5.61 	 \\ 
				&$^1E_{2g}  (\Val; \pi \ra \pis)$$^b$          &8.28	&8.31	&7.82$^c$&8.51	&				&$^3\Sigma_u^+ (\Val; \pi \ra \pis)$		& 4.10	&4.11	&3.67	&4.08	\\
				&$^3B_{1u} (\Val; \pi \ra \pis)$        		& 4.16	&4.22	&3.92	&4.32	&				&$^3\Delta_u 	(\Val; \pi \ra \pis)$		& 4.78	&4.82	&4.45	&4.78 	 \\
				&$^3E_{1u}(\Val; \pi \ra \pis)$	        		& 4.85	&4.89	&4.51	&4.92	&Diazomethane	&$^1A_2 (\Val; \pi \ra \pis)$   			& 3.14	&3.13	&2.92	&3.09 	 \\
				&$^3B_{2u} (\Val; \pi \ra \pis)$        		& 5.81	&5.85	&5.40	&5.51	&				&$^1B_1 (\mathrm{R}; \pi \ra 3s)$		& 5.54	&5.50	&5.30	&5.63 	 \\
Butadiene			&$^1B_u  (\Val; \pi \ra \pis)$	        		& 6.22	&6.76	&6.52	&6.68	&         		        &$^1A_1 (\Val; \pi \ra \pis)$   			& 5.90	&6.21	&5.92	&6.23 	 \\
				&$^1B_g (\mathrm{R}; \pi \ra 3s)$       	& 6.33	&6.49	&6.32	&6.44	&         		        &$^3A_2 (\Val; \pi \ra \pis)$   			& 2.79	&2.87	&2.67	&2.83 	 \\
         			&$^1A_g  (\Val; \pi \ra \pis)$$^b$      		& 6.50	&6.74	&6.30$^c$&6.70	&         		        &$^3A_1 (\Val; \pi \ra \pis)$   			& 4.05	&4.10	&3.88	&4.07 	 \\ 
				&$^1A_u (\mathrm{R}; \pi \ra 3p)$       	& 6.64	&6.74	&6.64	&6.84	&				&$^3B_1 (\mathrm{R}; \pi \ra 3s)$		& 5.35	&5.34	&5.15	&5.48 	 \\
				&$^1A_u (\mathrm{R}; \pi \ra 3p)$       	& 6.80	&6.95	&6.84	&7.01	&				&$^3A_1 (\mathrm{R}; \pi \ra 3p)$		& 6.82	&7.00	&6.76	&7.01 	 \\
				&$^1B_u (\mathrm{R}; \pi \ra 3p)$ 	        & 7.68	&7.60	&7.30	&7.45	&			& $^1A^{''}  [\mathrm{F}]  (\Val; \pi \ra \pis)$ 	& 0.71	&0.69	&0.52	&0.66 	 \\
				&$^3B_u (\Val; \pi \ra \pis)$	        		& 3.36	&3.40	&3.19	&3.40	&Formamide             &$^1A'' (\Val; n \ra \pis)$				& 5.65	&5.66	&5.45	&5.71	\\
				&$^3A_g (\Val; \pi \ra \pis)$	        		& 5.20	&5.32	&4.93	&5.30	&				&$^1A{'} (\mathrm{R}; n \ra 3s)$		& 6.77	&6.80	&6.64	&6.98 	 \\
				&$^3B_g (\mathrm{R}; \pi \ra 3s)$       	& 6.29	&6.44	&6.27	&6.38	&				&$^1A{'} (\mathrm{R}; n \ra 3p)$& \emph{{7.38}}  	&7.45	&7.32	&7.64 	 \\
Carbon trimer		&$^1\Delta_g   (\Val; n,n  \ra \pis,\pis)$	&5.22	&5.08	&4.85$^c$&5.25	&    		    		  &$^1A{'} (\Val; \pi \ra \pis)$	& \emph{{7.63}} 	&7.88	&7.13	&7.64 	 \\
				&$^1\Sigma^+_g (\Val; n,n  \ra \pis,\pis)$	&5.91	&5.82	&5.58$^d$&5.99	&                		&$^3A{''} (\Val; n \ra \pis)$				& 5.38	&5.36	&5.16	&5.38	\\
Cyanoacetylene	&$^1\Sigma^- 	(\Val; \pi \ra \pis)$   		& 5.80	&5.85	&5.47	&5.78	&               	 	&$^3A{'} (\Val; \pi \ra \pis)$			& 5.81	&5.88	&5.62	&5.90	\\					
				&$^1\Delta 	(\Val; \pi \ra \pis)$   		& 6.07	&6.13	&5.78	&6.10	&Furan			&$^1A_2 (\mathrm{R}; \pi \ra 3s)$		& 6.09 	&6.16	&6.04	&6.28 	 \\
				&$^3\Sigma^+	 (\Val; \pi \ra \pis)$   		& 4.44	&4.45	&4.04	&4.45	&				&$^1B_2 (\Val; \pi \ra \pis)$			& 6.37 	&6.59	&6.02	&6.20 	 \\
				&$^3\Delta 	(\Val; \pi \ra \pis)$ 		& 5.21	&5.21	&4.86	&5.19	&				&$^1A_1 (\Val; \pi \ra \pis)$			& 6.56 	&6.66	&6.10	&6.77 	 \\
				&$^1A'' [\mathrm{F}](\Val; \pi \ra \pis)$ 	& 3.54	&3.67	&3.47	&3.50	&				&$^1B_1  (\mathrm{R}; \pi \ra 3p)$ 		& 6.64	&6.79	&6.63	&6.71 	 \\
Cyanoformaldehyde	        &$^1A'' (\Val; n \ra \pis)$			& 3.81	&3.98	&3.67	&3.98	&				&$^1A_2  (\mathrm{R}; \pi \ra 3p)$		& 6.81 	&6.87	&6.77	&6.99 	 \\
				&$^1A'' (\Val; \pi \ra \pis)$				& 6.46	&6.79	&6.43	&6.44 	&				&$^1B_2  (\mathrm{R}; \pi \ra 3p)$		& 7.24    	&7.11 	&6.84	&7.01 	 \\
				&$^3A'' (\Val; n \ra \pis)$				& 3.44	&3.46	&3.25	&3.58 	&				&$^3B_2 (\Val; \pi \ra \pis)$			& 4.20    	&4.26	&4.01	&4.42	 \\
				&$^3A' (\Val; \pi \ra \pis)$				& 5.01	&5.25	&5.03	&5.35 	&				&$^3A_1 (\Val; \pi \ra \pis)$			& 5.46    	&5.50	&5.09	&5.60	 \\
Cyanogen			& $^1\Sigma_u^- (\Val; \pi \ra \pis)$		& 6.39	&6.40	&6.03	&6.32 	&				&$^3A_2 (\mathrm{R}; \pi \ra 3s)$		& 6.02    	&6.16	&5.99	&6.08	 \\
				& $^1\Delta_u (\Val; \pi \ra \pis)$ 		& 6.66 	&6.70	&6.35	&6.66 	&				&$^3B_1 (\mathrm{R}; \pi \ra 3p)$ 		& 6.59    	&6.76	&6.60	&6.68	 \\
				& $^3\Sigma_u^+ (\Val; \pi \ra \pis)$		& 4.91	&4.85	&4.46	&4.88 	&Glyoxal			&$^1A_u (\Val; n \ra \pis)$				& 2.88	&2.82	&2.51	&2.90	 \\
		& $^1\Sigma_u^-  [\mathrm{F}]  (\Val; \pi \ra \pis)$	& 5.05 	&5.07	&4.75	&4.97 	&				&$^1B_g (\Val; n \ra \pis)$				& 4.24	&4.21	&3.89$^c$&4.30	 \\
Cyclopentadiene	                &$^1B_2 (\Val; \pi \ra \pis)$	& 5.56	&5.96	&5.62	&5.65 	&				&$^1A_g (\Val; n,n  \ra \pis,\pis)$		& 5.61	&5.37   	&5.21   	&5.52	 \\
				&$^1A_2 (\mathrm{R}; \pi \ra 3s)$	 	& 5.78	&5.88	&5.78	&5.92 	&				&$^1B_g (\Val; n \ra \pis)$	 			& 6.57 	&6.52	&5.98$^c$&6.64	 \\
				&$^1B_1  (\mathrm{R}; \pi \ra 3p)$	 	& 6.41	&6.59	&6.44	&6.42	&				&$^1B_u (\mathrm{R}; n \ra 3p)$	        	& 7.71	&7.61	&7.34	&7.84	 \\
				&$^1A_2  (\mathrm{R}; \pi \ra 3p)$	 	& 6.46	&6.55	&6.46	&6.59	&				&$^3A_u (\Val; n \ra \pis)$				& 2.49	&2.41	&2.12	&2.49 	 \\
				&$^1B_2  (\mathrm{R}; \pi \ra 3p)$	 	& 6.56	&6.72	&6.56	&6.60	&				&$^3B_g (\Val; n \ra \pis)$				& 3.89	&3.90	&3.53	&3.99 	 \\
				&$^1A_1 (\Val; \pi \ra \pis)$$^b$& \emph{{6.52}}   	&6.63	&6.13	&6.75	&				&$^3B_u (\Val; \pi \ra \pis)$			& 5.15	&5.14	&4.91	&5.17 	 \\
				&$^3B_2 (\Val; \pi \ra \pis)$			& 3.31	&3.34	&3.09	&3.41	& 				&$^3A_g (\Val; \pi \ra \pis)$			& 6.30	&6.32	&6.02	&6.33 	 \\
				&$^3A_1 (\Val; \pi \ra \pis)$			& 5.11	&5.14	&4.78	&5.30	&Imidazole		&$^1A'' (\mathrm{R}; \pi \ra 3s)$		& 5.70 	&5.88	&5.66	&5.93 	 \\
				&$^3A_2 (\mathrm{R}; \pi \ra 3s)$	 	& 5.73	&5.91	&5.74	&5.73	&				&$^1A' (\mathrm{R}; \pi \ra 3p)$		& 6.41	&6.69	&6.45	&6.73 	 \\
				&$^3B_1 (\mathrm{R}; \pi \ra 3p)$	 	& 6.36	&6.56	&6.40	&6.40	&				&$^1A'' (\mathrm{R}; \pi \ra 3p)$		&6.50 	&6.57	&6.47	&6.83 	 \\
Cyclopropene	                &$^1B_1 (\Val; \si \ra \pis)$	        	& 6.68  	&6.86	&6.58	&6.80	&				&$^1A'' (\Val; n \ra \pis)$		        		& 6.71	&6.94	&6.57	&6.96 	 \\
				&$^1B_2 (\Val; \pi \ra \pis)$			& 6.79	&6.89	&6.47	&6.83	&				&$^1A' (\Val; \pi \ra \pis)$				& 6.86  	&6.88	&6.46	&7.00 	 \\ 
				&$^3B_2 (\Val; \pi \ra \pis)$			& 4.38	&4.47	&4.27	&4.51	&				&$^1A' (\mathrm{R}; n \ra 3s)$		    	&7.00 	&7.10	&6.91	&7.20 	 \\	 
				&$^3B_1 (\Val; \si \ra \pis)$	         	& 6.45	&6.56	&6.32	&6.52	&				&$^3A' (\Val; \pi \ra \pis)$				& 4.73	&4.78	&4.53	&4.86 	 \\
				&								&		&		&		&		&				&$^3A'' (\mathrm{R}; \pi \ra 3s)$		& 5.66	&5.86	&5.63	&5.91 	 \\
				&								&		&		&		&		&				&$^3A' (\Val; \pi \ra \pis)$				& 5.74	&5.85	&5.48	&5.91 	 \\
				&								&		&		&		&		&				&$^3A'' (\Val; n \ra \pis)$				& 6.31	&6.44	&6.10	&6.48 	 \\
ine
\end{tabular}
\vspace{-0.4cm}
\begin{flushleft}
$^a${All values are given in eV and computed with the {\AVTZ} basis. {The TBEs listed in italics are considered as unsafe.}}
$^b${Significant double excitation character;}
$^c${Level shift=0.4 a.u.;}
$^d${Level shift=0.5 a.u.}
\end{flushleft}
\end{table*}	

\begin{table*}[htp]
\caption{Comparisons between TBEs taken from Table 6 of Ref.~\citenum{Loo18a} and Table 11 of Ref.~\citenum{Loo20a} and VTEs computed at the 
 CASPT2(IPEA), CASPT2(NOIPEA), and (PC-)NEVPT2 values. $^a$} 
\label{Table-2}
\vspace{-0.4 cm}
\scriptsize
\begin{tabular}{llcccc|llcccc}
ine 
	Compound & State	& TBE & \multicolumn{2}{c}{CASPT2}    & NEVPT2 &Compound & State	& TBE & \multicolumn{2}{c}{CASPT2}    & NEVPT2 \\
	         &              &     & IPEA     & NOIPEA    &        &         &              &     & IPEA     & NOIPEA    &        \\
ine
Isobutene			&$^1B_1 (\mathrm{R}; \pi \ra 3s)$		& 6.46	&6.74	&6.59	&6.63   	& Pyridine			&$^1B_1 (\Val; n \ra \pis)$				& 4.95 	&5.15	&4.81	&5.15         \\
				&$^1A_1 (\mathrm{R}; \pi \ra 3p)$		& 7.01   	&7.32   	&7.14	&7.20   	& 				&$^1B_2 (\Val; \pi \ra \pis)$			& 5.14	&5.18	&4.76	&5.31        \\
				&$^3A_1 (\Val; \pi \ra \pis)$			& 4.53   	&4.59  	&4.41	&4.61   	& 				&$^1A_2 (\Val; n \ra \pis)$				& 5.40	&5.46	&5.03	&5.29         \\
Ketene                  	&$^1A_2 (\Val; \pi \ra \pis)$			& 3.86	&3.92	&3.70	&3.93   	& 				&$^1A_1 (\Val; \pi \ra \pis)$			& 6.62	&6.92	&6.27	&6.69        \\
				&$^1B_1 (\mathrm{R}; \pi \ra 3s)$		& 6.01	&5.99	&5.79	&6.09   	& 				&$^1A_1 (\mathrm{R}; n \ra 3s)$		& 6.76	&6.90	&6.67	&6.99        \\
				&$^1A_2 (\mathrm{R}; \pi \ra 3p)$		& 7.18	&7.25	&7.05	&7.28   	& 				&$^1A_2 (\mathrm{R}; \pi \ra 3s)$		& 6.82	&7.08	&6.87	&6.86        \\
				&$^3A_2(\Val; \pi \ra \pis)$			& 3.77  	&3.81	&3.59	&3.80   	& 				&$^1B_2 (\Val; \pi \ra \pis)$	&  \emph{{7.40}} 	&7.92	&7.67	&7.83        \\
				&$^3A_1(\Val; \pi \ra \pis)$			& 5.61  	&5.65	&5.43	&5.65   	& 				&$^1B_1 (\mathrm{R}; \pi \ra 3p)$		& 7.38	&7.70	&7.51	&7.45        \\
				&$^3B_1 (\mathrm{R}; \pi \ra 3s)$ 		& 5.79	&5.79	&5.60	&5.89   	& 				&$^1A_1 (\Val; \pi \ra \pis)$			& 7.39	&7.66	&6.63	&6.97         \\
				&$^3A_2 (\mathrm{R}; \pi \ra 3p)$		& 7.12	&7.22	&7.01	&7.24   	& 				&$^3A_1 (\Val; \pi \ra \pis)$			& 4.30	&4.40	&4.06	&4.60        \\
			& $^1A^{''}  [\mathrm{F}]  (\Val; \pi \ra \pis)$ 	& 1.00	&1.05	&0.88	&1.02   	& 				&$^3B_1 (\Val; n \ra \pis)$				& 4.46	&4.48	&4.21	&4.58         \\
Methylenecyclopropene         &$^1B_2 (\Val; \pi \ra \pis)$		& 4.28	&4.40	&4.12	&4.37 	& 				&$^3B_2 (\Val; \pi \ra \pis)$			& 4.79	&4.86	&4.53	&4.88 	\\
				&$^1B_1 (\mathrm{R}; \pi \ra 3s)$		& 5.44	&5.57	&5.44	&5.49 	& 				&$^3A_1 (\Val; \pi \ra \pis)$			& 5.04	&5.09	&4.63	&5.19 	\\
				&$^1A_2 (\mathrm{R}; \pi \ra 3p)$		& 5.96	&6.09	&5.97	&6.00  	& 				&$^3A_2 (\Val; n \ra \pis)$				& 5.36	&5.33	&4.96	&5.33 	\\
				&$^1A_1(\Val; \pi \ra \pis)$	& \emph{{6.12}}     &6.26	&6.16	&6.36 	& 				&$^3B_2 (\Val; \pi \ra \pis)$			& 6.24	&6.40	&5.99	&6.29 	\\
				&$^3B_2 (\Val; \pi \ra \pis)$			& 3.49	&3.57	&3.34	&3.66  	& Pyrimidine		&$^1B_1 (\Val; n \ra \pis)$				& 4.44	&4.44	&4.07	&4.55 	\\
				&$^3A_1 (\Val; \pi \ra \pis)$			& 4.74	&4.82	&4.58	&4.87 	& 				&$^1A_2 (\Val; n \ra \pis)$				& 4.85	&4.80	&4.36	&4.84  	\\
Nitrosomethane	&$^1A'' (\Val; n \ra \pis)$		       		&1.96	&1.84	&1.60	&1.91 	& 				&$^1B_2 (\Val; \pi \ra \pis)$			& 5.38	&5.40	&5.01	&5.53 	\\
				&$^1A' (\Val; n,n  \ra \pis,\pis)$			&4.76	&4.69	&4.67	&4.73 	& 				&$^1A_2 (\Val; n \ra \pis)$				& 5.92	&5.92	&5.32	&6.02  	\\
				&$^1A' (\mathrm{R}; n  \ra 3s)$			&6.29	&6.32	&6.07	&6.38 	& 				&$^1B_1 (\Val; n \ra \pis)$				& 6.26	&6.31	&5.65	&6.40 	\\
         			&$^3A'' (\Val; n \ra \pis)$		        		& 1.16	&1.00	&0.75	&1.08 	& 				&$^1B_2  (\mathrm{R}; n \ra 3s)$	        & 6.70	&6.85	&6.50	&6.77 	\\
				&$^3A{'} (\Val; \pi \ra \pis)$		        & 5.60	&5.52	&5.37	&5.54  	& 				&$^1A_1 (\Val; \pi \ra \pis)$			& 6.88	&7.23	&6.88	&7.11 	\\
			& $^1A^{''}  [\mathrm{F}] (\Val; n \ra \pis)$ 	        & 1.67	&1.55	&1.32	&1.62 	& 				&$^3B_1 (\Val; n \ra \pis)$				& 4.09	&4.05	&3.67	&4.17	\\
Propynal			& $^1A'' (\Val; n \ra \pis)$				& 3.80 	&3.92	&3.64	&3.95  	& 				&$^3A_1 (\Val; \pi \ra \pis)$	&  \emph{{4.51}}  	&4.56	&4.25	&4.67	\\
				&$^1A'' (\Val; \pi \ra \pis)$				& 5.54	&5.82	&5.49	&5.50 	& 				&$^3A_2 (\Val; n \ra \pis)$				& 4.66	&4.63	&4.16	&4.72	\\
				&$^3A'' (\Val; n \ra \pis)$				& 3.47	&3.48	&3.26	&3.59 	& 				&$^3B_2 (\Val; \pi \ra \pis)$			& 4.96	&4.98	&4.67	&5.01	\\
				&$^3A' (\Val; \pi \ra \pis)$				& 4.47	&4.59	&4.30	&4.63 	& Pyrrole 			&$^1A_2 (\mathrm{R}; \pi \ra 3s)$		& 5.24	&5.44	&5.23	&5.51	\\
Pyrazine			&$^1B_{3u}  (\Val; n \ra \pis)$			& 4.15	&4.09	&3.66	&4.17 	& 				&$^1B_1 (\mathrm{R}; \pi \ra 3p)$		& 6.00 	&6.26	&6.07	&6.32	\\
				&$^1A_{u}  (\Val; n \ra \pis)$			& 4.98	&4.76	&4.26	&4.77 	& 				&$^1A_2 (\mathrm{R}; \pi \ra 3p)$		& 6.00	&6.16	&6.02	&6.44	\\
				&$^1B_{2u}  (\Val; \pi \ra \pis)$			& 5.02	&5.13	&4.65	&5.32  	& 				&$^1B_2 (\Val; (\pi \ra \pis)$			& 6.26 	&6.62	&6.36	&6.48	\\
				&$^1B_{2g}  (\Val; n \ra \pis)$			& 5.71	&5.68	&5.27	&5.88 	& 				&$^1A_1 (\Val; \pi \ra \pis)$			& 6.30 	&6.41	&5.84	&6.53	\\
				&$^1A_{g}  (\mathrm{R}; n \ra 3s)$		& 6.65	&6.66	&6.27	&6.70  	& 				&$^1B_2 (\mathrm{R}; \pi \ra 3p)$		& 6.83	&6.75	&6.11	&6.62	\\
				&$^1B_{1g}  (\Val; n \ra \pis)$			& 6.74	&6.61	&6.07	&6.75 	& 				&$^3B_2 (\Val; \pi \ra \pis)$			& 4.51	&4.57	&4.30	&4.74	\\
				&$^1B_{1u}  (\Val; \pi \ra \pis)$			& 6.88	&7.14	&6.72	&6.81 	& 				&$^3A_2 (\mathrm{R}; \pi \ra 3s)$		& 5.21	&5.41	&5.21	&5.49	\\
				&$^1B_{1g}  (\mathrm{R}; \pi \ra 3s)$	& 7.21	&7.41	&7.27	&7.33 	& 				&$^3A_1 (\Val; \pi \ra \pis)$			& 5.45	&5.50	&5.04	&5.56	\\
				&$^1B_{2u}  (\mathrm{R}; n \ra 3p)$		& 7.24	&7.34	&6.93	&7.25 	& 				&$^3B_1 (\mathrm{R}; \pi \ra 3p)$		& 5.91	&6.22	&6.03	&6.28	\\
				&$^1B_{1u}  (\mathrm{R}; n \ra 3p)$		& 7.44	&7.55	&7.08	&7.42 	& Streptocyanine-C1     	&$^1B_2 (\Val; \pi \ra \pis)$		& 7.13	&7.17	&6.76	&7.13	\\
				&$^1B_{1u}  (\Val; \pi \ra \pis)$	&  \emph{{7.98}}  	&8.59	&7.96	&8.25  	& 				&$^3B_2 (\Val; \pi \ra \pis)$	        		& 5.52	&5.49	&5.22	&5.52	\\
				&$^3B_{3u}  (\Val; n \ra \pis)$			& 3.59	&3.49	&3.08	&3.56 	& Tetrazine		&$^1B_{3u}  (\Val; n \ra \pis)$			& 2.47	&2.31	&1.91	&2.35	\\
				&$^3B_{1u}  (\Val; \pi \ra \pis)$			& 4.35	&4.44	&4.15	&4.57  	& 				&$^1A_{u}  (\Val; n \ra \pis)$			& 3.69	&3.49	&3.00	&3.58	\\
				&$^3B_{2u}  (\Val; (\pi \ra \pis)$			& 4.39	&4.44	&4.09	&4.42 	& 			&$^1A_{g}  (\Val; n,n \ra \pis, \pis)$$^b$&\emph{{4.61}} &4.69   	&4.48   	&4.61	\\
				&$^3A_{u}  (\Val; n \ra \pis)$		        & 4.93	&4.73	&4.21	&4.75 	& 				&$^1B_{1g}  (\Val; n \ra \pis)$			& 4.93	&4.83	&4.33	&4.95	\\
				&$^3B_{2g}  (\Val; n \ra \pis)$			& 5.08	&5.04	&4.66	&5.21 	& 				&$^1B_{2u}  (\Val; \pi \ra \pis)$			& 5.21	&5.31	&4.84	&5.56	\\
				&$^3B_{1u}  (\Val; \pi \ra \pis)$			& 5.28	&5.29	&4.92	&5.35 	& 				&$^1B_{2g}  (\Val; n \ra \pis)$			& 5.45	&5.38	&4.90	&5.63	\\
Pyridazine			&$^1B_1 (\Val; n \ra \pis)$				& 3.83	&3.74	&3.36	&3.80 	& 				&$^1A_{u}  (\Val; n \ra \pis)$			& 5.53	&5.51	&4.92	&5.62	\\
				&$^1A_2 (\Val; n \ra \pis)$				& 4.37   	&4.29   	&3.87	&4.40  	& 			&$^1B_{3g} (\Val; n,n \ra \pis, \pis)$$^b$	&\emph{{6.15}}&5.85   	&5.22   	&6.15	\\
				&$^1A_1 (\Val; \pi \ra \pis)$			& 5.26 	&5.34	&4.87	&5.58 	& 				&$^1B_{2g}  (\Val; n \ra \pis)$			& 6.12	&5.96	&5.18	&6.13	\\
				&$^1A_2 (\Val; n \ra \pis)$				& 5.72	&5.73	&5.19	&5.88  	& 				&$^1B_{1g}  (\Val; n \ra \pis)$			& 6.91	&6.59$^c$&5.89$^c$&6.76	\\
				&$^1B_2  (\mathrm{R}; n \ra 3s)$		& 6.17	&6.18	&5.90	&6.21 	& 				&$^3B_{3u}  (\Val; n \ra \pis)$			& 1.85	&1.70	&1.31	&1.73	\\
				&$^1B_1 (\Val; n \ra \pis)$				& 6.37	&6.50	&5.94	&6.64 	& 				&$^3A_{u}  (\Val; n \ra \pis)$			& 3.45	&3.26	&2.78	&3.36	\\
				&$^1B_2 (\Val; \pi \ra \pis)$			& 6.75	&7.21	&6.74	&7.10	&				&$^3B_{1g}  (\Val; n \ra \pis)$			& 4.20	&4.10	&3.62	&4.24	\\
				&$^3B_1 (\Val; n \ra \pis)$				& 3.19	&3.08	&2.72	&3.13 	& 				&$^3B_{1u}  (\Val; \pi \ra \pis)$	&\emph{{4.49}}  	&4.55	&4.29	&4.70	\\
				&$^3A_2 (\Val; n \ra \pis)$				& 4.11	&4.01	&3.59	&4.14 	& 				&$^3B_{2u}  (\Val; \pi \ra \pis)$			& 4.52	&4.55	&4.20	&4.58	\\
				&$^3B_2 (\Val; \pi \ra \pis)$	&  \emph{{4.34}}	&4.40	&4.12	&4.49	&				&$^3B_{2g}  (\Val; n \ra \pis)$			& 5.04	&5.02	&4.53	&5.27	\\
				&$^3A_1 (\Val; \pi \ra \pis)$			& 4.82	&4.87	&4.48	&4.94 	& 				&$^3A_{u}  (\Val; n \ra \pis)$			& 5.11	&5.07	&4.44	&5.13	\\
 				&								&		&		&		&		&			&$^3B_{3g}  (\Val; n,n \ra \pis, \pis)$$^b$&\emph{{5.51}}&5.39	&4.86  	&5.51	\\
				&								&		&		&		&		&				&$^3B_{1u}  (\Val; \pi \ra \pis)$			& 5.42	&5.46	&5.08	&5.56	\\
ine
\end{tabular}	
\vspace{-0.4cm}
\begin{flushleft}
$^a${ {See caption of Table} \ref{Table-1} {for details; $^b$ For these three genuine doubly-excited states, the PC-NEVPT2 values are the current TBEs.}}
$^c${The level shift is set to $0.4$ a.u.}
\end{flushleft}
\end{table*}

\begin{table*}[htp]
\caption{Comparisons between TBEs taken from Table 6 of Ref.~\citenum{Loo18a} and Table 11 of Ref.~\citenum{Loo20a} and VTEs computed at the 
 CASPT2(IPEA), CASPT2(NOIPEA), and  (PC-)NEVPT2 values.$^a$} 
\label{Table-3}
\vspace{-0.4 cm}
\scriptsize
\begin{tabular}{llcccc|llcccc}
ine 
	Compound & State	& TBE & \multicolumn{2}{c}{CASPT2}    & NEVPT2 &Compound & State	& TBE & \multicolumn{2}{c}{CASPT2}    & NEVPT2 \\
	         &              &     & IPEA     & NOIPEA    &        &         &              &     & IPEA     & NOIPEA    &        \\
ine
Thioacetone		        &$^1A_2 (\Val; n \ra \pis)$			& 2.53 	&2.58	&2.33	&2.55 	&Thiopropynal		        &$^1A''  (\Val; n \ra \pis)$			& 2.03	&2.05	&1.84	&2.05 \\
				&$^1B_2 (\mathrm{R}; n \ra 4s)$		& 5.56 	&5.60	&5.48	&5.72 	&				&$^3A''   (\Val; n \ra \pis)$				& 1.80	&1.81	&1.60	&1.81 \\
				&$^1A_1 (\Val; \pi \ra \pis)$			& 5.88	&6.42	&5.98	&6.24 	&Triazine			&$^1A_1'' (\Val; n \ra \pis)$			& 4.72	&4.62	&3.90	&4.61 \\
				&$^1B_2 (\mathrm{R}; n \ra 4p)$	        	& 6.51	&6.51	&6.40	&6.62 	&				&$^1A_2'' (\Val; n \ra \pis)$			& 4.75	&4.77	&4.39	&4.89 \\
				&$^1A_1 (\mathrm{R}; n \ra 4p)$		& 6.61	&6.66	&6.41	&6.52 	&				&$^1E'' (\Val; n \ra \pis)$				& 4.78	&4.76	&4.14	&4.88 \\
				&$^3A_2 (\Val; n \ra \pis)$				& 2.33 	&2.34	&2.09	&2.32 	&				&$^1A_2' (\Val; \pi \ra \pis)$	        		& 5.75	&5.76	&5.32	&5.95 \\
				&$^3A_1 (\Val; \pi \ra \pis)$			& 3.45	&3.48	&3.29	&3.48 	&				&$^1A_1' (\Val; \pi \ra \pis)$			& 7.24	&7.43	&6.89	&7.30 \\
Thiophene		        &$^1A_1 (\Val; \pi \ra \pis)$		& 5.64	&5.84	&5.21	&5.84 	&				&$^1E' (\mathrm{R}; n \ra 3s)$			& 7.32	&7.48	&7.15	&7.45 \\ 
				&$^1B_2 (\Val; \pi \ra \pis)$			& 5.98	&6.35	&5.89	&6.10 	&				&$^1E'' (\Val; n \ra \pis)$				& 7.78	&7.75   	&7.04   	&7.98 \\
				&$^1A_2 (\mathrm{R}; \pi \ra 3s)$		& 6.14	&6.28	&6.07	&6.20 	&				&$^1E' (\Val; \pi \ra \pis)$				& 7.94	&8.65	&7.70	&8.34 \\
				&$^1B_1 (\mathrm{R}; \pi \ra 3p)$		& 6.14	&6.21	&5.90	&6.19 	&				&$^3A_2'' (\Val; n \ra \pis)$			& 4.33	&4.37	&3.99	&4.51 \\
				&$^1A_2 (\mathrm{R}; \pi \ra 3p)$		& 6.21	&6.32	&5.98	&6.40 	&				&$^3E'' (\Val; n \ra \pis)$				& 4.51	&4.47	&3.88	&4.61 \\
				&$^1B_1 (\mathrm{R}; \pi \ra 3s)$		& 6.49	&6.57	&6.28	&6.71 	&				&$^3A_1'' (\Val; n \ra \pis)$			& 4.73	&4.70	&3.94	&4.71 \\
				&$^1B_2 (\mathrm{R}; \pi \ra 3p)$		& 7.29 	&7.29	&7.03	&7.25 	&				&$^3A_1' (\Val; \pi \ra \pis)$			& 4.85	&4.88	&4.55	&5.05 \\
				&$^1A_1 (\Val; \pi \ra \pis)$	&  \emph{{7.31}}  	&7.62	&6.85	&7.39  	&				&$^3E' (\Val; \pi \ra \pis)$				& 5.59	&5.62	&5.20	&5.73 \\
				&$^3B_2 (\Val; \pi \ra \pis)$			& 3.92	&3.98	&3.71	&4.13 	&				&$^3A_2' (\Val; (\pi \ra \pis)$			& 6.62	&6.62	&6.12	&6.36 \\
				&$^3A_1 (\Val; \pi \ra \pis)$			& 4.76	&4.85	&4.39	&4.84 	&\\
				&$^3B_1 (\mathrm{R}; \pi \ra 3p)$		& 5.93	&5.97	&5.64	&5.98 	&\\
				&$^3A_2 (\mathrm{R}; \pi \ra 3s)$		& 6.08	&6.22	&6.01	&6.14 	&\\
ine
\end{tabular}	
\vspace{-0.3cm}
\begin{flushleft}
$^a${ {See caption of Table} \ref{Table-1} {for details}.}
\end{flushleft}
\end{table*}	

\subsection{Molecules with three non-hydrogen atoms}

We have estimated VTEs in eight molecules containing three non-H atoms, namely, acetaldehyde, carbon trimer, cyclopropene, diazomethane, formamide, ketene, nitrosomethane, and streptocyanine-C1. Reference FCI (or high-level CC) 
values with various basis sets, as well as literature data can be found in our previous works. \cite{Loo18a,Loo19c} The choice of active spaces and state-averaging procedure are detailed in Tables S1--S7 in the SI, except for carbon trimer 
where it was described elsewhere. \cite{Loo19c}

\emph{Acetaldehyde.} The VTEs computed for the lowest $n \ra \pis$ singlet and triplet transitions of acetaldehyde obtained with NEVPT2 at $4.39$ and $4.00$ eV, respectively, slightly exceed older estimates obtained at the same level of theory 
($4.29$ and $3.97$ eV), \cite{Ang05b} and both are fitting the TBEs of $4.31$ and $3.97$ eV, obtained on the basis of FCI calculations.\cite{Loo18a} Likewise CASPT2(IPEA) estimates are also accurate at $4.35$ and $3.94$ eV, whereas the absence of 
IPEA shift produces underestimated values ($4.13$ and $3.71$ eV).

\emph{Carbon trimer.} This original linear molecule is of interest because it presents two ``pure'' low-lying doubly-excited states for which FCI-quality TBEs are available. As one can see from the data listed in Table \ref{Table-1}, NEVPT2 provides values
extremely close to FCI, whereas the absence of IPEA shift in CASPT2 yields very large underestimations.

\emph{Cyclopropene.} The lowest two singlet $\sigma \ra \pis$ and $\pi \ra \pis$ transitions of cyclopropene, of respective $B_1$ and $B_2$ symmetries,  are very close in energy, \cite{Sch08,Loo18a} an effect reproduced by all 
second-order perturbation  theories considered here. While both CASPT2(IPEA) and NEVPT2 deliver the same ordering as CCSDT, \cite{Loo18a} CASPT2(NOIPEA) swaps the two states and it significantly underestimates the VTE associated with the 
$\pi \ra \pis$ excitation. All VTEs obtained with CASPT2(IPEA) are within ca. $0.1$ eV of the TBEs, except for the singlet $\sigma \ra \pis$ transition for which the error is slightly larger (i.e., within $0.2$ eV). Switching off the IPEA shift downshifts all 
CASPT2 estimates from $0.2$ eV up to $0.4$ eV, leading to larger errors.

\emph{Diazomethane.} The $\pi \ra \pis$ transition energies of both singlet and triplet symmetries of diazomethane obtained from CASPT2(IPEA) are very close to the TBEs, except for the valence $^1A_1 (\pi \ra \pis)$ transition 
for which an error of $0.3$ eV is obtained. A similar observation can be made for NEVPT2. VTEs for Rydberg excitations are also nicely reproduced by both methods with errors around $0.1$ eV except for the $^3A_1 (\mathrm{R}; \pi \ra 3p)$ transition 
for which the error increases to $0.2$ eV. CASPT2(NOIPEA) underestimates all the TBEs, apart from the valence $^1A_1 (\pi \ra \pis)$ transition which is overestimated upon inclusion of the IPEA correction. For the eight transitions considered 
the MAEs are $0.09$, $0.15$, and $0.11$ eV for CASPT2(IPEA), CASPT2(NOIPEA), and NEVPT2, respectively.

\emph{Formamide.} The lowest singlet and triplet $n \ra \pis$ transitions of formamide obtained with CASPT2(IPEA) are within $0.02$ eV of our TBEs, whereas the CASPT2(NOIPEA) values are too small. NEVPT2 is also very accurate for these 
two transitions. For the three higher singlet states of $A^{'}$ symmetry, the presence of a strong state mixing between the valence and Rydberg transitions was observed in previous CC studies by us \cite{Loo18a} and the Szalay's group.
\cite{Kan17} This mixing led to a problematic assignment of these states, as the lowest Rydberg state has a larger oscillator strength than the valence $(\pi \ra \pis)$ state, which is counterintuitive. Our SA-CASSCF reference 
wave function shows that such a mixing is weak and the assignment of the states is straightforward, highlighting a clear advantage of multi-reference methods in this case. The oscillator strengths are also consistent with the electronic 
nature of the states, the valence $(\pi \ra \pis)$ state having the largest transition dipole moment ($3.2$ D) compared to the Rydberg states ($0.9$ and $1.6$ D). Overall, the results obtained with CASPT2(IPEA) are consistent with previous 
TBEs \cite{Loo18a}, although we do not consider these transitions in the statistics below due to the state-mixing issue.

\emph{Ketene.} For the eight ESs of ketene listed in Table \ref{Table-2}, the MAEs are $0.05$, $0.16$, and $0.07$ eV  for  CASPT2(IPEA), CASPT2(NOIPEA), and NEVPT2, respectively. As in the previous systems, CASPT2(NOIPEA) 
underestimates systematically all the TBEs. Ketene was previously investigated at the MS-CASPT2/6-31+G(d) level by the Morokuma group, \cite{Xia13} who reported VTEs of $3.72$, $5.97$,
$3.62$, $5.42$, and $5.69$ eV for the $^1A_2$, $^1B_1$, $^3A_2$, $^3A_1$, and $^3B_1$ excitations, respectively, all values being slightly below the TBEs. Despite the different basis sets used, one also notes a reasonable match
between these previous MS-CASPT2 values and the present results that rely on rather comparable active spaces.

\emph{Nitrosomethane.} Both the lowest-lying singlet and triplet transitions of nitrosomethane are of  $n \ra \pis$ character and the NEVPT2 VTEs are in very good agreement with the TBEs, whereas CASPT2 provides
too small values.  More interestingly, the second singlet ES has a pure double electronic excitation of $n, n \ra \pis, \pis$ nature, and a FCI value of $4.76$ eV could be obtained. 
\cite{Loo19c} In contrast to single-reference methods that miserably fail for this transition except when quadruples are included, \cite{Loo19c} all three second-order multireference perturbative methods tested here deliver very
accurate results with a slight underestimation by less than $0.1$ eV.

\emph{Streptocyanine-C1.} The shortest cyanine is an interesting case as its transition cannot be accurately described by TD-DFT. \cite{Sen11,Leg15} At the CASPT2 level, our values are comparable to the
ones obtained by Send \textit{et al.}, \cite{Sen11} both with and without IPEA shift. The VTEs of the valence singlet and triplet $\pi \ra \pis$  obtained from CASPT2(IPEA) and NEVPT2 levels of 
theory are also close from one another, with the NEVPT2 values being in perfect agreement with the TBEs.\cite{Loo18a} The absence of IPEA correction again yields significantly underestimated values.

\subsection{Molecules with four non-hydrogen atoms}

For these 15 systems, as well as for the larger systems treated below, the description of the corresponding active spaces can be found in our earlier contribution. \cite{Loo20a} 

\emph{Acetone and thioacetone.}  Previous works reporting CASPT2 and NEVPT2 VTEs exist for acetone. \cite{Roo96,Ang05b,Sch08,Sil10c,Pas12,Loo20a} The present CASPT2(IPEA) results are in excellent agreement 
with both their counterparts of Ref.~\citenum{Pas12} and the TBEs, except for the  $^1A_2 (n \ra 3p)$ transition that appears significantly too high.  Interestingly, the VTE associated with this transition is nicely reproduced by 
CASPT2(NOIPEA).  For the lowest $n \ra \pis$ excitations of both spin symmetries, CASPT2(IPEA) provides VTEs in very close agreement with the ones reported by Roos and co-workers.  \cite{Roo96} The NEVPT2 estimates are 
accurate for the valence ESs, but too large for the Rydberg ESs.  In thioacetone, the valence  $^1A_1 (\pi \ra \pis)$ transition is considerably overshot by both CASPT2(IPEA) and NEVPT2. For all other ESs, both 
approaches deliver very satisfying accuracy, with no absolute error exceeding $0.2$ eV. Again, turning off the IPEA shift is detrimental for all transitions except the $^1A_1 (\pi \ra \pis)$ one.

\emph{Acrolein.}  The {most} comprehensive previous CASPT2 study of acrolein is likely the work of Aquilante, Barone, and Roos, \cite{Aqu03} which reports many transitions. For the nine ESs listed in Table \ref{Table-1},
the lack of IPEA shift clearly yields significant underestimations (except for the ES having a significant share of double excitation character) whereas both CASPT2(IPEA) and NEVPT2 values 
are trustworthy, the latter leading a small average deviations as compared to the TBEs.  However, one specific challenging ES is the second $^1A'' (n \ra \pis)$ valence excitation, for which the VTEs produced by all 
approaches seem inconsistent. One should note, however, that this transition has a significant contribution from the doubly-excited configurations, \cite{Loo20a} making its CCSDT-based TBE likely less accurate.

\emph{Butadiene.}  The relative VTEs of the optically bright 1$^1B_u$ and dark 2$^1A_g$ ESs have certainly been the topic of many theoretical studies, given both the experimental interest and the
mixed single/double excitation character of the latter ES.  \cite{Ser93,Ost01,Sah06,Wat12,Ise12,Ise13,Sch17,Shu17,Sok17,Chi18,Cop18,Tra19,Loo20a} The first very reliable estimates are likely due to Watson and 
Chan, \cite{Wat12} who showed that the bright ESs should be slightly lower in energy. Our $6.22$ and $6.50$ eV TBEs, based on CCSDTQ and FCI results, respectively, \cite{Loo19c,Loo20a} follow this trend and can likely be 
considered as reliable.  As can be seen, none of the three tested multiconfigurational approaches accurately reproduced the gap between these two ESs, CASPT2(NOIPEA) being very poor, confirming that butadiene 
remains a particularly stringent test. For the other ESs, the trends noted above are conserved, i.e., both NEVPT2 and CASPT2(IPEA) provide reliable valence/overestimated Rydberg VTEs,
whereas CASPT2(NOIPEA) tends to be more reliable for the Rydberg ESs.

\emph{Cyanoacetylene, cyanogen, and diacetylene.} For these three closely related linear molecules, the TBEs are based on CCSDTQ, and the difference between CC3, CCSDT, and CCSDTQ
values are totally negligible, \cite{Loo20a} strongly hinting that the TBEs are highly trustworthy. On average, both CASPT2(IPEA) and NEVPT2 are rather competitive  
with deviations no larger than $\pm 0.10$ eV except for the fluorescence of cyanoacetylene with CASPT2(IPEA). In contrast, CASPT2(NOIPEA) systematically produces underestimated VTEs for 
all 13 ESs of this series.

\emph{Cyanoformaldehyde, propynal, and thiopropynal.}  These three molecules possess a conjugated carbonyl group and are of $C_s$ symmetry. All the considered ESs have a
strong  single-excitation character \cite{Loo20a}. The VTEs of the lowest singlet transitions, of $n \ra \pis$ nature, are almost identical between NEVPT2 and CASPT2(IPEA), with
absolute values slightly exceeding the TBEs. Surprisingly, the methodological differences are larger for the corresponding triplet ES in both cyanoformaldehyde and propynal, for which CASPT2(IPEA) is
extremely accurate. In contrast, for the $A' (\pi \ra \pis)$ transition of cyanoformaldehyde, not applying an IPEA shift seems beneficial.

\emph{Cyclopropenone, cyclopropenethione, and methylenecyclopropene.}  These three molecules are characterized by a three-membered $sp^2$ carbon cycle conjugated
to an external $\pi$ bond. There are previous multiconfigurational studies for these three compounds. \cite{Mer96,Ser02,Liu14b,Bud17,Loo20a} For cyclopropenone, the VTEs
of the singlet and triplet valence $n \ra \pis$ transitions of $A_2$ symmetry and singlet $n \ra 3s$ Rydberg transition of $B_2$ symmetry
obtained with CASPT2(IPEA) nicely fit the reference CCSDTQ-based values. \cite{Loo20a} In contrast, the VTEs of the singlet and triplet $n \ra \pis$ transitions of $B_1$ 
symmetry are underestimated by all second-oder approaches assessed here. For cyclopropenethione, the 
CASPT2 values reported by  Serrano-Andr\'{e}s et al. \cite{Ser02} are typically in between the CASPT2(IPEA) and CASPT2(NOIPEA)  values listed in Table \ref{Table-1}.
CASPT2(IPEA) outperforms the two other methods reported in this Table, except for the $^1B_2 (n \ra 3s)$ and $^1A_1 (\pi \ra \pis)$ transitions. For methylenecyclopropene, the
current CASPT2(NOIPEA) results fit well with those reported by Roos and co-workers. \cite{Mer96} Pinpointing the most suitable level of theory is challenging from the
data of Table \ref{Table-2} as the MAEs are similar for methylenecyclopropene: $0.11$, $0.10$, and $0.12$ eV for CASPT2(IPEA), CASPT2(NOIPEA), and NEVPT2, respectively.

\emph{Glyoxal.} In glyoxal, one needs to separate the lowest $^1A_g$ ES, of pure double $(n,n  \ra \pis,\pis)$ nature, from all other considered ESs that are strongly dominated
by single excitations. For the latter set of transitions, both CASPT2(IPEA) and NEVPT2 typically provide accurate estimates, whereas the lack of IPEA correction gives significant underestimations,
even for the Rydberg transition considered herein.  For the specific $^1A_g$ ES, our TBE is based on FCI, \cite{Loo19c,Loo20a} and the NEVPT2 VTE is compatible
with this estimate, whereas both CASPT2 approaches deliver too small transition energies. Interestingly the SAC-CI method in its so-called general-$R$ form provides a VTE
of $5.66$ eV for this state, \cite{Sah06} actually outperforming the three multiconfigurational methods considered in the present study.

\emph{Isobutene.} Similarly to other cases, both CASPT2(IPEA) and NEVPT2 significantly overshoot the singlet TBEs of the Rydberg ESs, whereas for the lowest triplet of valence nature, 
these two approaches yield results within $0.1$ eV of the reference value.

\subsection{Five-membered cycles}

\emph{Cyclopentadiene.}  There are several previous CASPT2 \cite{Ser93b,Sch08,Sil10c,Bud17} and NEVPT2 \cite{Loo20a} studies of VTEs in cyclopentadiene. None of the ES treated  here, except the 
$^1A_1 (\pi \ra \pis)$ transition, has a significant contribution from the doubly-excited determinants. However, even for this transition, one notes in Table \ref{Table-1} a reasonable agreement between the CASPT2(IPEA), 
NEVPT2,  and the CCSDT-based TBEs (that all slightly exceed earlier estimates \cite{Ser93b,Sch08,Sil10c}) as well as the most recent experimental value we are aware of. \cite{McD85} For all other singlet
transitions, including the lowest $^1B_2 (\pi \ra \pis)$ excitation, CASPT2(NOIPEA) provides VTEs closer to the TBEs than CASPT2(IPEA). For the triplet ESs considered here, one finds again the usual
pattern with valence (Rydberg) VTEs more accurate when the IPEA shift is turned on (off). Considering the 10 ESs, one gets MAEs of $0.14$, $0.13$, and $0.12$ eV for CASPT2(IPEA),  CASPT2(NOIPEA), and 
NEVPT2, respectively.

\emph{Furan.}  Unsurprisingly, furan was also investigated with several multireference methods. \cite{Ser93b,Pas06b,Sch08,Sil10c,Li10c,Bud17}  We consider only  ESs possessing a highly-dominant
single excitation character. For both the singlet and triplet valence $\pi \ra \pis$ transitions of $B_2$ and $A_1$ symmetries, the present CASPT2(NOIPEA) values are reasonably close to those
determined by Roos' group, \cite{Ser93b} whereas blueshifts are found when including the IPEA shift.  \cite{Sch08,Sil10c} An interesting case is the  $^1B_2 (\pi \ra 3p)$ excitation for which our CCSDT-based TBE of
$7.24$ eV is larger than the estimates given in most previous theoretical studies (ca.~$6.5$--$6.9$ eV). \cite{Ser93b,Pas06b,Sch08,Sil10c,Li10c,Hol15} For this particular ES, the present NEVPT2 and CASPT2(IPEA)
estimates are also above $7$ eV, indicating that this ES is very sensitive to the employed methodological details. For the 10 transitions reported in Table \ref{Table-1}, the MAEs are
$0.10$, $0.23$, and $0.18$ eV for CASPT2(IPEA),  CASPT2(NOIPEA), and NEVPT2, respectively, that is, the relative ranking of the methods differs from the one obtained for cyclopentadiene.

\emph{Imidazole.} The most detailed previous CASPT2 investigation is due to  Serrano-Andr\'{e}s and co-workers, \cite{Ser96b} and the VTEs we report here with
CASPT2(NOIPEA) for the Rydberg transitions, as well as the singlet and triplet valence $\pi \ra \pis$ transitions of $A^{'}$~symmetry are quite close to these previous
values.  The quality of these VTEs, as compared to the TBEs, \cite{Loo20a}  are rather contrasted. Again, we found that CASPT2(NOIPEA) is more effective for the Rydberg
than the valence transitions.

\emph{Pyrrole.} The VTEs of pyrrole were extensively studied using second-order perturbation theory \cite{Ser93b,Roo02,Sch08,Sil10c,Hol15,Hei19,Loo20a}. We underline that 
our TBEs \cite{Loo20a} are in very nice agreement with XMS-CASPT2 results obtained by the Gonz\'{a}lez group, \cite{Hei19} while the present CASPT2(NOIPEA) values are larger than the 
1993 VTEs proposed by Serrano-Andr\'{e}s et al. \cite{Ser93b} For the singlet transitions, it is difficult to highlight clear trends for both CASPT2 approaches, whereas for the triplet ESs the 
usual tendencies are found. Interestingly NEVPT2 systematically overestimates the VTEs for all considered ESs, except for the highest-lying  singlet Rydberg state.

\emph{Thiophene.} The optical properties of the most popular sulfur-bearing cycle have been investigated both experimentally,  \cite{DiL72, Gyo13} and with multiconfigurational approaches. \cite{Ser93c,Pas07,Loo20a} 
Again, all ESs listed in Table \ref{Table-3} are of dominant single-excitation character though the two singlet $A_1$ transitions have non-negligible contributions from higher excitations.  CASPT2(NOIPEA)  systematically yields too small
VTEs, with an average deviation of $-0.24$ eV.  Turning on the IPEA shift somehow overcorrects, as CASPT2(IPEA) produces systematic overestimations by an average of $+0.13$ eV. NEVPT2
delivers a more balanced treatment, with no deviation exceeding a quarter of an eV and a MAE of $0.11$ eV.

\subsection{Six-membered cycles}

\emph{Benzene.} Previous multiconfigurational studies, \cite{Lor95b,Sch08,Sil10c,Loo19c,Sha19,Loo20a} as well as refined experimental measurements  \cite{Doe69,Hir91} are again available for benzene.
We find that our CASPT2(NOIPEA) data for the singlet Rydberg transitions are systematically higher in energy than the previous CASPT2 results of Roos and co-workers, \cite{Lor95b}  whereas our TBEs
reasonably fit published  RASPT2 data. \cite{Sha19} The trends obtained by analyzing the results of Table \ref{Table-1} are rather clear: (i) CASPT2(IPEA) outperforms both CASPT2(NOIPEA) 
and NEVPT2 for estimating the valence $\pi \ra \pis$~transitions; (ii)  CASPT2(NOIPEA) is accurate for Rydberg transitions but leads to too small values for the valence ESs.

\emph{Pyrazine.} There are previous multireference estimates of the VTEs of pyrazine,\cite{Ful92,Web99,Sch08,Woy10,Sil10c,Sau11,Loo19c,Loo20a} and all ESs considered can be described as single 
excitations. Indeed, the ES with the largest contribution from higher excitations, the $^1B_{1g}$ $n \ra \pis$ transition, has a single character of $84.2$\%\ according to CC3. \cite{Loo20a} From the data listed in Table \ref{Table-2}, it
is clear that CASPT2(NOIPEA) is not very accurate, but it is not straightforward to determine if CASPT2(IPEA) is superior to NEVPT2 (or the opposite). An interesting specificity is that the lowest $^1A_{u}$
ES seems rather challenging for both methods.

\emph{Pyridazine.} As for previous systems, the present VTEs are consistent with previous CASPT2 studies including or not IPEA shifts. \cite{Ful92,Fis00,Sch08,Sil10c} It is also noteworthy that our  
CASPT2(NOIPEA) data are in close proximity  with the very recent electron energy-loss  experiments, \cite{Lin19} but, as always, such comparison can only be viewed as qualitative. Considering
all pyridazine ESs of Table \ref{Table-2}, one obtains MAEs of $0.11$, $0.38$, and $0.14$ eV, for  CASPT2(IPEA),  CASPT2(NOIPEA), and NEVPT2, respectively. The absence of IPEA shift is systematically
detrimental, even for the Rydberg $^1B_2$ transition.

\emph{Pyridine.} The ESs of pyridine were the subject of multiconfigurational \cite{Ful92,Lor95,Sch08,Sil10c,Sau11,Loo20a} and refined experimental \cite{Wal90,Lin16} studies. We found a good correlation 
between the current CASPT2(NOIPEA) data and the early estimates of the Roos group. \cite{Lor95}  Again, the VTEs of the Rydberg transitions, e.g., the singlet $n/\pi \ra 3s$ transitions of $A_1$ and 
$A_2$ symmetry, obtained with CASPT2(NOIPEA) are in good agreement with the TBEs, whereas for the valence transitions, both CASPT2(IPEA) and NEVPT2 provide mode accurate VTEs.  Those
two latter methods provide similar MAEs  (ca.~$0.20$ eV) for pyridine.

\emph{Pyrimidine.} Earlier CASPT2 and NEVPT2 estimates are available for this azabenzene. \cite{Ful92,Ser97b,Fis03b,Sch08,Sil10c,Sau11,Loo20a} The results of Table \ref{Table-2}
reveal the excellent accuracy obtained with CASPT2(IPEA).  There is only one ES, namely, $^1A_1 (\pi \ra \pis)$, for which a deviation larger than $0.15$ eV could be detected.
For this state, of strong single-excitation character, \cite{Loo20a} not applying an IPEA shift is beneficial.  NEVPT2 also provides very reasonable estimates except for a small overestimation
trend. Nevertheless, NEVPT2 outperforms both CASPT2 schemes for the Rydberg $^1B_2  (n \ra 3s)$ transition, which is an unusual outcome for the panel of compounds
treated here.

\emph{Tetrazine.} In this intensively studied, \cite{Rub99,Dev08,Sch08,Ang09,Sil10c,Loo19c,Loo20a} highly symmetric compound, one can clearly distinguish ESs of single-excitation character 
to those that are pure double excitations, namely, $^1A_g$, $^1B_{3g}$, and $^3B_{3g}$. \cite{Loo19c,Loo20a} For these three particular states, our TBEs were previously classified as \emph{``unsafe''}, 
\cite{Loo20a} as FCI estimates are beyond computational reach, so that indisputable TBEs are not available. For the $^1A_g$ transition, the CASPT2(IPEA) and NEVPT2 values are very 
close from another, but differences of $0.30$ and $0.12$ eV are obtained for the singlet and triplet $B_{3g}$ transitions, respectively, justifying their \emph{``unsafe''} status. Focussing
on the single-excitation transitions, one notes very good performances of both CASPT2(IPEA) and NEVPT2 with respective MAEs of $0.11$ and $0.12$ eV.

\emph{Triazine.} Finally, for the $D_{3h}$ azabenzene, the present CASPT2(NOIPEA) results are in line with the early investigations of Roos and co-workers \cite{Ful92} 
and Serrano-Andr\'{e}s and co-workers \cite{Oli05}  whereas the CASPT2(IPEA) data fit reasonably well the CASPT2 values determined later by Thiel's group. \cite{Sch08,Sil10c} 
When comparing to the TBEs, it is pretty obvious that CASPT2(IPEA) is the most adequate approach for most ESs, the absence of IPEA shift leading to systematic 
underestimations of the VTEs, whereas NEVPT2 produces slightly too large transition energies.

\subsection{Statistical analysis}

\begin{figure*}[htp]
\centering
 \includegraphics[width=.8\linewidth]{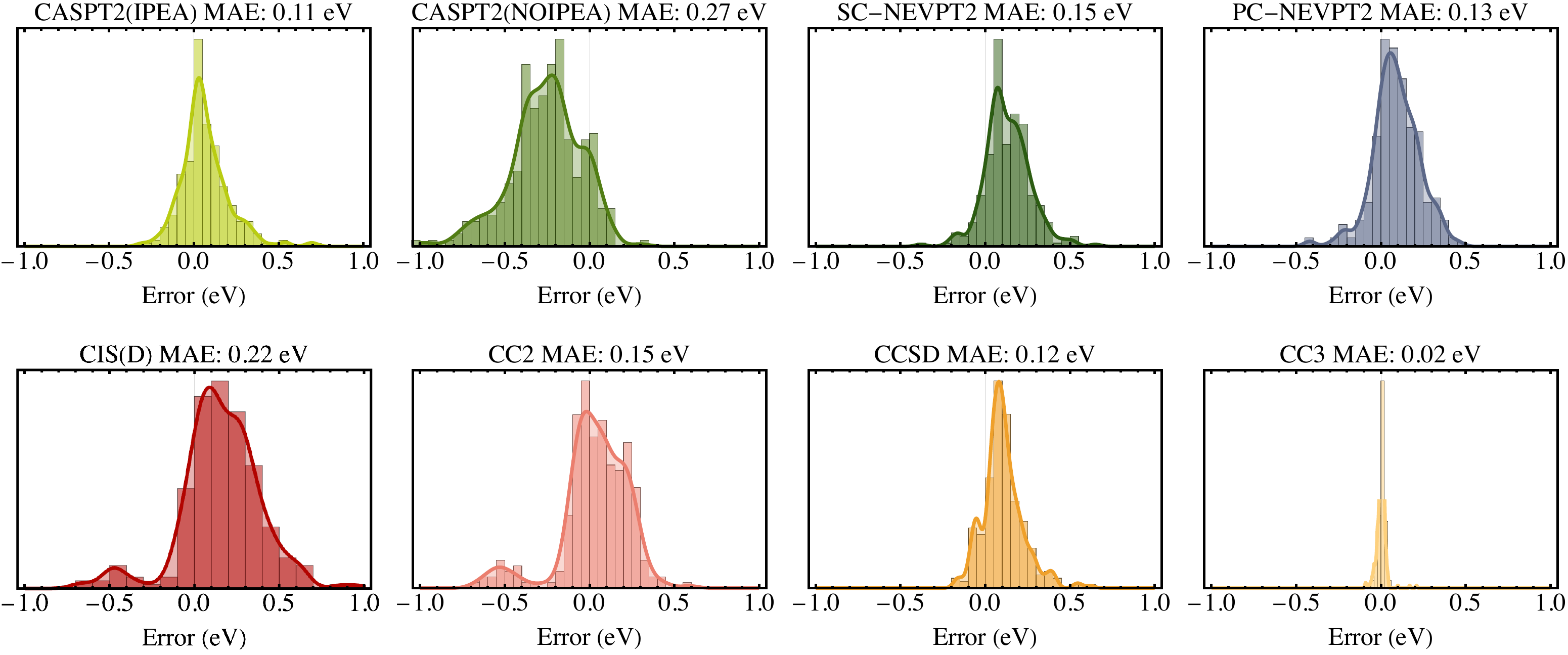}
  \caption{Histograms of the errors obtained for the four multiconfigurational second-order perturbation theory methods and comparisons with a selection of single-reference methods.}
   \label{Fig-2}
\end{figure*}

To perform our statistical analysis, we only take into account the VTEs qualified as \emph{``safe''} in our previous studies. \cite{Loo20a,Ver21} Therefore, we removed, for example, transitions of double excitation character
for which FCI energies could not be computed, the strongly mixed ESs in formamide, and the other troublemakers {for which the TBEs have been noted in italics in Tables} \ref{Table-1}--\ref{Table-3}.  
The corresponding statistical indicators are gathered in Table \ref{Table-4} and the histogram representation of the spread of the errors can also be found in Figure \ref{Fig-2}.

\begin{table}[htp]
\footnotesize
\vspace{-0.3 cm}
\caption{Statistical analysis, taking the TBEs as reference, for the four multiconfigurational approaches.   At the bottom
of the Table, we provide similar data extracted from the QUEST database for single-reference methods. \cite{Ver21} 
{Count is the number of ESs considered in the statistics.}
All {error} values are in eV. }
\label{Table-4}
\begin{tabular}{l|ccccccc}
ine
Method 			& Count 	& MSE 	&MAE 	&RMSE 	&SDE 	&Max($+$)&Max($-$)	\\
ine
SA-CASSCF		&265		&$0.13	$&$0.48$	&$0.60$	&$0.57$	&$2.14$	&$-1.18$\\
CASPT2(IPEA)		&265		&$0.06	$&$0.11$	&$0.16$	&$0.14$	&$0.71$	&$-0.32$\\
CASPT2(NOIPEA)	&265		&$-0.26	$&$0.27$	&$0.33$	&$0.21$	&$0.30$	&$-1.02$\\
SC-NEVPT2		&265		&$0.13	$&$0.15$	&$0.19$	&$0.13$	&$0.65$	&$-0.38$\\
PC-NEVPT2		&265		&$0.09	$&$0.13$	&$0.16$	&$0.13$	&$0.46$	&$-0.42$\\
ine
CIS(D)			&257		&$0.16$	&$0.22$	&$0.28$	&$0.23$	&$0.96$	&$-0.69$\\
ADC(2)			&252		&$0.00$	&$0.14$	&$0.19$	&$0.18$	&$0.64$	&$-0.73$\\
CC2				&258		&$0.03$	&$0.15$	&$0.20$	&$0.19$	&$0.59$	&$-0.68$\\
SCS-CC2			&258		&$0.15$	&$0.17$	&$0.21$	&$0.15$	&$0.76$	&$-0.92$\\
CCSD			&259		&$0.10$	&$0.12$	&$0.16$	&$0.05$	&$0.62$	&$-0.17$\\
CC3				&262		&$0.00$	&$0.02$	&$0.03$	&$0.03$	&$0.21$	&$-0.09$\\
ine
 \end{tabular}
 \end{table}

In line with the analysis performed for individual molecules, it turns out that CASPT2 delivers slightly too large VTEs when the IPEA shift  is turned on ($+0.06$ eV) but large underestimations when the IPEA shift is set to zero ($-0.26$ eV). The 
MAE obtained with CASPT2(IPEA) is $0.11$ eV, which can certainly be viewed as acceptable for many practical applications. It, nevertheless, remains far from chemical accuracy ($1$ kcal.mol$^{-1}$ or $0.04$ eV). The dispersion of the 
errors, as measured by the SDE, is also significantly larger when the IPEA shift is not applied ($0.21$ eV \emph{vs} $0.14$ eV), which clearly shows that the IPEA shift must be applied in practical CASPT2 calculations of VTEs in organic molecules.  
This observation is consistent with the findings of the Gonz\'{a}lez group, who concluded that, for triple-$\zeta$ basis sets, the IPEA shift was beneficial. \cite{Zob17} Interestingly, PC-NEVPT2, which is free of the IPEA dilemma, delivers almost 
the same performance as CASPT2(IPEA), with slightly larger overestimations of the VTEs, but slightly smaller spread of the errors and maximal deviations.  In other words, the present benchmark does not reveal any significant difference in 
terms of accuracy between these two multiconfigurational approaches.  SC-NEVPT2 is found to be only slightly worse than PC-NEVPT2. In contrast (and as expected), SA-CASSCF is unsatisfying with a MAE as large as $0.48$ eV 
and a large dispersion as well.  

At the bottom of Table \ref{Table-4}, we provide the results obtained for the same set of compounds and ESs with a selection of popular single-reference methods. For the sake of consistency, these values have been straightforwardly extracted from the 
QUEST database, \cite{Ver21} but are in line with many other benchmarks. \cite{Hat05c,Sch08,Wat13,Kan14,Har14,Kan17} As can be seen ADC(2), CC2, and CCSD do deliver average deviations of the same order of magnitude as both CASPT2(IPEA) 
and PC-NEVPT2, but with a slightly larger dispersion of the errors for ADC(2) and CC2. The latter effect can be attenuated by using the spin-scaled variant (SCS) of CC2, but at the cost of inducing a nearly systematic overestimation, as evidenced by 
the large positive MSE.  This means that one should be very cautious in using CASPT2 data as reference to benchmark the quality of ADC(2)'s or CC2's VTEs in well-behaved transitions. Indeed, the superiority of the former model over the two others 
is not perfectly clear.

\begin{figure*}[htp]
\centering
 \includegraphics[width=.8\linewidth]{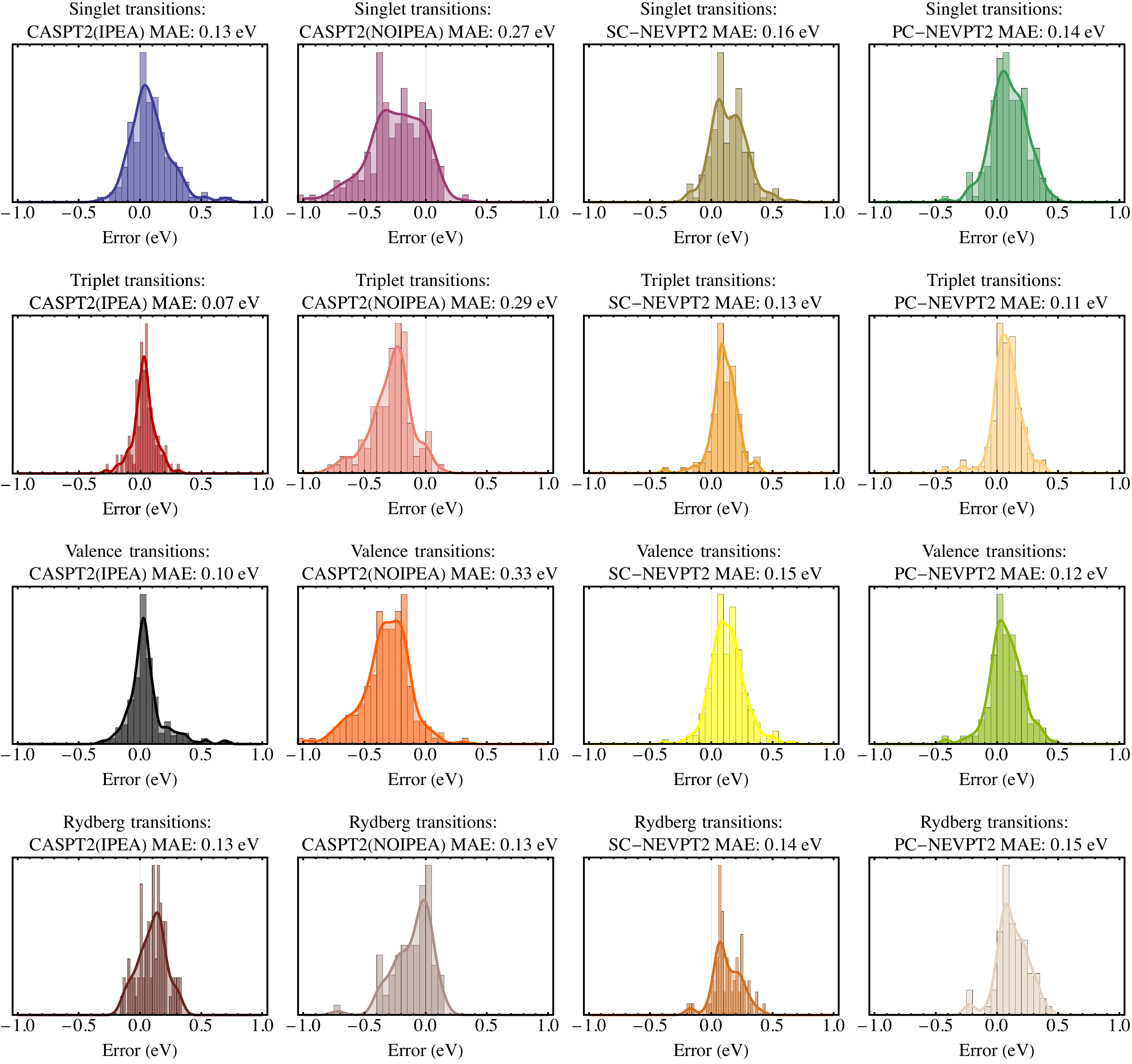}
  \caption{Histograms of the errors for various subsets obtained with the four multiconfigurational second-order perturbation theory methods.}
   \label{Fig-3}
\end{figure*}

In Table \ref{Table-5}, we provide the MAEs determined for various subsets of transitions, whereas Figure \ref{Fig-3}  provides a graphical comparison for selected families of transitions. Clearly CASPT2(NOIPEA) delivers reasonable estimates 
for the Rydberg transitions only, which is fully consistent with the analyses made for individual systems above.  SA-CASSCF appears more accurate for the triplet, valence, and $n \ra \pis$ transitions than for the singlet,
Rydberg, and $\pi \ra \pis$ counterparts, but the deviations remain very large in all cases. Although one can notice small differences for various subsets, i.e., slightly improved performances for triplet, valence, and $n \ra \pis$ ESs as compared 
to the singlet, Rydberg, and $\pi \ra \pis$ subsets, it is a pleasant outcome that both CASPT2(IPEA) and the two NEVPT2 variants deliver rather equivalent and satisfactory levels of accuracy for all single-excitation subsets.  For the systems 
considered here, rather similar trends were observed for ADC(2) and CC2, though these two models are more effective for the $ n \ra \pis$ than the $\pi \ra \pis$ excitations. Finally, the transitions with a dominant contribution from the 
doubly-excited states are more accurately modeled by NEVPT2 than CASPT2, while single-reference approaches are simply unable to describe these transitions.

\begin{table*}[htp]
\footnotesize
\caption{MAE determined for several subsets computed at various levels of theory. {See caption of Table} \ref{Table-4} {for details.}}
\label{Table-5}
\begin{tabular}{l|ccccccc}
ine
Method 			&Singlet	& Triplet	& Valence 	&Rydberg	&$n \ra \pis$	& $\pi \ra \pis$ 	&Double\\
ine
SA-CASSCF		&$0.56$	&$0.34$	&$0.45$	&$0.54$	&$0.44$	&$0.46$	&$0.42$\\
CASPT2(IPEA)		&$0.13$	&$0.07$	&$0.10$	&$0.13$	&$0.08$	&$0.12$	&$0.14$\\
CASPT2(NOIPEA)	&$0.27$	&$0.29$	&$0.33$	&$0.13$	&$0.44$	&$0.26$	&$0.30$\\
SC-NEVPT2		&$0.16$	&$0.13$	&$0.15$	&$0.14$	&$0.12$	&$0.17$	&$0.04$\\
PC-NEVPT2		&$0.14$	&$0.11$	&$0.12$	&$0.15$	&$0.10$	&$0.13$	&$0.06$\\
ine
CIS(D)			&$0.21$	&$0.24$	&$0.25$	&$0.16$	&$0.21$	&$0.27$	&\\
ADC(2)			&$0.15$	&$0.13$	&$0.13$	&$0.16$	&$0.09$	&$0.16$	&\\
CC2				&$0.16$	&$0.14$	&$0.14$	&$0.17$	&$0.07$	&$0.19$	&\\
SCS-CC2			&$0.16$	&$0.19$	&$0.20$	&$0.10$	&$0.22$	&$0.19$	&\\
CCSD			&$0.15$	&$0.09$	&$0.14$	&$0.09$	&$0.18$	&$0.11$	&\\
CC3				&$0.02$	&$0.01$	&$0.02$	&$0.01$	&$0.01$	&$0.02$	&\\
ine
 \end{tabular}
 \end{table*}
 
Besides, by looking into the evolution of the errors as the size of the compounds increases, one notes a deterioration of the MAEs for CASPT2(IPEA) and both NEVPT2 schemes with, e.g., MAEs of $0.08$, $0.12$, and $0.15$ eV for the 3, 4, and 5-6 non-hydrogen atom molecules for PC-NEVPT2. The CASPT2(NOIPEA) deviations follow a similar pattern, but with larger errors, with a MAE of 
$0.21$ eV for the 3 non-hydrogen atom molecules, but $0.32$ eV for the 5-6 non-hydrogen atom compounds.

We can compare the deviations obtained here to the ones reported in Ref.~\citenum{Sch13b} using Thiel's CC3/TZVP values as reference. For 121 singlet (72 triplets) valence ESs, Schapiro \textit{et al.} obtained MAEs of 
$0.23$ ($0.23$), $0.28$ ($0.25$), and $0.21$ ($0.14$) eV for SC-NEVPT2, PC-NEVPT2, and SS-CASPT2(IPEA), respectively, that is, deviations significantly larger than the present ones. However, we note that their key 
conclusions stating that the three approaches deliver similar results, and that CASPT2 is slightly more accurate for the triplet ESs are fully consistent with the present findings. Analyzing the data of this earlier work, it appears 
that differences with the current results can be mostly ascribed to the $\pi \ra \pis$ transitions (that are more accurate here quite possibly due to the use of a larger diffuse-containing basis set), and likely to the total absence of 
Rydberg and doubly-excited transitions in Thiel's set. In Ref.~\citenum{Sch13b}, the absence of IPEA correction produced very large underestimations of the CASPT2 VTEs (MSE of $-0.48$ eV for singlet transitions). This is 
again consistent with the trend reported here, but with a more dramatic error likely due to the particular focus on valence transitions. Finally, we note that for SA-CASSCF, much larger errors have been reported when using the 
122 Thiel CC3/TZVP reference values, \cite{Hel19} with a MAE of $0.98$ eV. Nonetheless, the fact that SA-CASSCF is more trustworthy for $ n \ra \pis$ than $\pi \ra \pis$ transitions is consistent with this study. \cite{Hel19}

\section{Conclusions}

We have computed VTEs of more than 250 ESs in 35 small to medium size organic molecules containing three to six non-hydrogen atoms using four different perturbative approaches and compared these VTEs with previous TBEs
established with highly-accurate methods.\cite{Loo20a,Ver21} 

Besides containing a wealth of data for individual molecules and specific transitions, this study allows to extract general trends.  In this regard, the global statistical analyses reveal that the overall performance of CASPT2(NOIPEA) 
is not up to the mark, except maybe for Rydberg transitions. Indeed, not applying an IPEA shift typically leads to strong underestimations of the transition energies together with a significant spread of the errors. The same statistical 
analyses indicate that both CASPT2(IPEA) and PC-NEVPT2 do yield small overestimations of the VTEs and provide more accurate estimates for the different subsets of ESs, with global MAEs of $0.11$ and $0.13$ eV, and SDEs of 
$0.14$ and $0.13$ eV, respectively.  Neglecting dynamic correlation, as in SA-CASSCF, is clearly very detrimental, whereas using the SC instead of the PC variant of NEVPT2 only slightly deteriorates the results. As compared to 
single-reference models, one notes that CASPT2(IPEA) and PC-NEVPT2 deliver similar accuracies as ADC(2), CC2, and CCSD, but with smaller dispersion than the first two methods. As a consequence, one should likely be very 
cautious when comparing ADC(2) and CASPT2 to attribute the error mainly to one of the two methods. It is also noteworthy that the CC3 VTEs remain much more accurate than the multiconfigurational ones. 

Of course, the present study has some limitations. First, for each of the ESs treated herein, one can certainly define a larger active space leading to a more accurate VTE. Yet, we trust that our active spaces are very reasonable, that is,
they would be aligned with typical choices made by CASSCF experts. {The choice of active space is made such as to describe mainly the static correlation associated with the electronic excitations at ground-state geometries. Describing excited states away from the equilibrium geometry, e.g. in the region of conical intersections in photochemistry, may require adapting the choice of active orbitals to describe the electronic reorganization along the reaction path.} Second, though we tested the use of IPEA ($0.25$ a.u.) or its complete neglect (as typically done in the literature), one could also consider searching for an 
optimal value. This aspect was previously treated by others \cite{Zob17} and it was demonstrated to be fairly challenging to define a general value applicable across various basis sets and systems. Third (and we wish to particularly 
stress this point), the present investigation was focussed on ESs with a dominant single-excitation character, meaning that the present conclusions regarding the relative performances of CASSCF, CASPT2, NEVPT2, and the various 
other single-reference approaches are obviously limited to this specific class of ESs.

{Finally, we provide in the SI a Table with the formal scaling costs of various single and multi-reference methods. Although the actual computational cost will be highly-dependent on the actual problem under investigation, this 
might be helpful to choose a specific approach. Of course, for the multi-reference approaches, the size of the active space is a critical parameter. PC-NEVPT2 is about one and a half to three times faster than CASPT2 with the implementation of the code used in this study.
}

\section*{Acknowledgements}
RS and DJ are indebted to the R\'egion des Pays de la Loire program for support in the framework of the Opt-Basis grant. PFL thanks the European Research Council (ERC) under the European Union's Horizon 2020 
research and innovation programme (Grant agreement No.~863481) for financial support. This research used computational resources of (i) the GENCI-TGCC (Grant No.~2019-A0060801738);  (ii) CALMIP under 
allocation 2020-18005 (Toulouse); (iii) CCIPL (\emph{Centre de Calcul Intensif des Pays de Loire});  (iv) a local Troy cluster and (v) HPC resources from ArronaxPlus  (grant ANR-11-EQPX-0004 funded by the 
French National Agency for Research). 

\section*{Supporting Information Available}
Description of active spaces for molecules not treated previously. Corrections and additions of previous data. CASSCF and SC-NEVPT2 transitions energies.
The Supporting Information is available free of charge at https://pubs.acs.org/doi/10.1021/{doi}.

\bibliography{biblio-new}

\end{document}